%% file: main.tex
\documentclass[lettersize,journal]{IEEEtran}
\IEEEoverridecommandlockouts

\usepackage{amsmath,amsfonts}
 
\usepackage{amssymb}

\usepackage{pifont} 

\RequirePackage[figuresright]{rotating}

\usepackage{pifont} 
\usepackage{amssymb}
\usepackage{bbding} 
\usepackage{pifont}
\newcommand{\cmark}{\textcolor{green!80!black}{\ding{51}}}
\newcommand{\xmark}{\textcolor{red}{\ding{55}}}

\usepackage{makecell}
\usepackage{colortbl} 
\newcommand{\hlhref}[2]{\href{#1}{\textcolor{blue}{\underline{#2}}}}

\usepackage{xcolor}
\definecolor{ao(english)}{rgb}{0.0, 0.5, 0.0}

\usepackage{tikz}
\usetikzlibrary{arrows.meta}
\usetikzlibrary{automata,positioning}

\usepackage{threeparttable}

\renewcommand*{\arraystretch}{1.5}%
\definecolor{tabred}{RGB}{230,36,0}%
\definecolor{tabgreen}{RGB}{0,116,21}%
\definecolor{taborange}{RGB}{250,124,30}%
\definecolor{tabbrown}{RGB}{171,70,0}%
\definecolor{tabyellow}{RGB}{251,253,169}%
\newcommand*{\vcorr}{%
  \vadjust{\vspace{-\dp\csname @arstrutbox\endcsname}}%
  \global\let\vcorr\relax
}%

\newcommand{\Low}{\textcolor{red}{\textbf{Low}}}
\newcommand{\Medium}{\textcolor{orange}{\textbf{Medium}}}
\newcommand{\High}{\textcolor{green!60!black}{\textbf{High}}}
\newcommand{\VeryHigh}{\textcolor{red!75!black}{\textbf{Very High}}}
\newcommand{\Variable}{\textcolor{blue!70!black}{\textbf{Variable}}}
\newcommand{\LowMed}{\textcolor{orange!80!black}{\textbf{Low--Medium}}}
\newcommand{\MedHigh}{\textcolor{orange!90!black}{\textbf{Medium--High}}}
\newcommand{\RiskLow}[1]{\textcolor{green!60!black}{\textbf{#1}}}  
\newcommand{\RiskMed}[1]{\textcolor{orange!90!black}{\textbf{#1}}} 
\newcommand{\RiskHigh}[1]{\textcolor{red!75!black}{\textbf{#1}}}    

\usepackage{mathtools}

\usepackage{lscape}
\usepackage[figuresright]{rotating}
\usepackage[graphicx]{realboxes}
\usepackage{adjustbox}
\usepackage{caption}

\usepackage{subfigure}

\usepackage{colortbl} 
\usepackage{xfrac}

\usepackage{float}

\usepackage{fbox} 
\usepackage{fancybox}
\usepackage{framed}

\usepackage{multirow}

\def\BibTeX{{\rm B\kern-.05em{\sc i\kern-.025em b}\kern-.08em
    T\kern-.1667em\lower.7ex\hbox{E}\kern-.125emX}}
    
\usepackage{CJKutf8}  
\usepackage{pgfplots}
\usepackage{filecontents}

\usepackage{hyperref}

\usepackage{tabularx,makecell, caption}
\usepackage{enumitem} 

\newcolumntype{L}{>{\arraybackslash}X}

\usepackage{multicol}
\usepackage{listings}
\usepackage{blindtext}
\usepackage{etoolbox,xstring,mfirstuc,textcase}
\usepackage{listings}

\usepackage{subfiles}
\usepackage[most]{tcolorbox}
\usepackage[scaled=0.95]{inconsolata}

\usepackage{xcolor}
\usepackage{siunitx}
\usepackage{comment}
\usepackage{indentfirst} 
\usepackage{framed} 

\usepackage[font=small,labelfont=bf,tableposition=top]{caption}
\usepackage{booktabs}
\usepackage{dashbox}

\usepackage{longtable}
\usepackage{ulem}

\usepackage{listings}
\lstdefinelanguage{Solidity}{
  keywords={contract, function, returns, uint, string, public, view, pure, if, else, require, emit, event, mapping, address, memory, storage, return, modifier, bool, true, false, struct, for, while, break, continue, delete, new, this, payable, revert},
  keywordstyle=\color{blue}\bfseries,
  ndkeywords={uint256, int, uint8, override},
  ndkeywordstyle=\color{teal}\bfseries,
  identifierstyle=\color{black},
  sensitive=true,
  comment=[l]{//},
  morecomment=[s]{/*}{*/},
  commentstyle=\color{gray}\ttfamily,
  stringstyle=\color{orange}\ttfamily,
  morestring=[b]",
  morestring=[b]'
}

\definecolor{codegray}{gray}{0.97}  
\definecolor{jsonkey}{rgb}{0.13, 0.13, 1}  
\definecolor{jsonstring}{rgb}{0.8, 0.2, 0.2}  

\lstdefinelanguage{json}{
  basicstyle=\ttfamily\footnotesize,
  backgroundcolor=\color{codegray},
  frame=single,
  showstringspaces=false,
  breaklines=true,
  captionpos=b,
  literate=
   *{0}{{{\color{black}0}}}{1}
    {1}{{{\color{black}1}}}{1}
    {2}{{{\color{black}2}}}{1}
    {3}{{{\color{black}3}}}{1}
    {4}{{{\color{black}4}}}{1}
    {5}{{{\color{black}5}}}{1}
    {6}{{{\color{black}6}}}{1}
    {7}{{{\color{black}7}}}{1}
    {8}{{{\color{black}8}}}{1}
    {9}{{{\color{black}9}}}{1}
    {:}{{{\color{black}:}}}{1}
    {,}{{{\color{black},}}}{1}
    {"}{{{\color{jsonstring}"}}}{1}
}

\usepackage{url}

\newcolumntype{Y}{>{\raggedright\arraybackslash}X}

\usepackage{color}
\usepackage{algorithm,algpseudocode} 

\begin{document}
\title{Understanding NFTs from EIP Standards}


\author{Minfeng Qi$^{1}$, Qin Wang$^{2,3}$, Guangsheng Yu$^{3}$,  Ruiqiang Li$^{4}$, Victor Zhou$^{5}$\thanks{$^\star$Victor Zhou is under EIP editor group, and also affiliated with Namefi. }, Shiping Chen$^{2}$
\\ 
\smallskip
\textit{$^1$City University of Macau}   $|$ \textit{$^2$CSIRO Data61} $|$  \textit{$^3$Univeristy of Technology Sydney} 
\\
 \textit{$^4$Univeristy of Wollongong}  $|$ \textit{$^5$Ethereum Foundation} 
}


\maketitle

\begin{abstract}
We argue that the technical foundations of non-fungible tokens (NFTs) remain inadequately understood. Prior research has focused on market dynamics, user behavior, and isolated security incidents, yet systematic analysis of the standards underpinning NFT functionality is largely absent.

We present the first study of NFTs through the lens of Ethereum Improvement Proposals (EIPs). We conduct a large-scale empirical analysis of 191 NFT-related EIPs and 10K+ Ethereum Magicians discussions (as of July, 2025). We integrate multi-dimensional analyses including the automated parsing of Solidity interfaces, graph-based modeling of inheritance structures, contributor profiling, and mining of community discussion data. We distinguish foundational from emerging standards, expose poor cross-version interoperability, and show that growing functional complexity heightens security risks.

\end{abstract}

\begin{IEEEkeywords}
NFT, EIPs, ERC, Blockchain
\end{IEEEkeywords}

\section{Introduction}
\noindent\textbf{NFTs are foundation of the web3 ecosystem.}
Non-fungible tokens (NFTs)~\cite{wang2021non} have evolved from a niche technological novelty into a foundational component of the Web3 ecosystem~\cite{wang2022exploring}. NFTs represent verifiable ownership of unique digital or physical assets on the blockchain~\cite{wang2023nfts}, enabling a wide range of applications across digital art, gaming, virtual real estate, identity verification, and decentralized finance (DeFi). This innovation has spurred a multi-billion-dollar market. Industry projections~\cite{tbrc2023nft} suggest that the global NFT market may expand from \$60.01 billion in 2025 to as much as \$247.41 billion by 2035, reflecting a compound annual growth rate (CAGR) of 41.9\%. Within this landscape, collectibles constitute the largest segment, accounting for 32.6\% of market activity, encompassing digital artworks, music, and video clips~\cite{coinlaw2025nft}.

NFT also attracts the engagement of Web3 users. In 2025, the decentralized application (dApp) ecosystem experienced a 485\% increase in unique active wallets (UAWs), reaching an average of 24.6 million UAWs per day by the end of the year~\cite{gherghelas2025dapp}. Notably, approximately 2.1 million of these daily wallets interacted with blockchain-based games, accounting for 30\% of all NFT-related activities~\cite{gherghelas2024state}. 

\smallskip
\noindent\textbf{What defines NFTs?} 
Despite their economic and cultural significance, NFTs do not exist in isolation; they are governed by a technical backbone comprising Ethereum Improvement Proposals (EIPs). An EIP is a general design document that outlines new features or processes for the Ethereum ecosystem. Within this broader set, Ethereum Request for Comments (ERCs) constitute a specific category of EIPs\footnote{We use the terms EIP and ERC interchangeably throughout this work.} focused on application-level standards for smart contracts. ERCs define how tokens, wallets, and decentralized applications should behave to ensure interoperability across the network.

NFTs are defined by ERCs, which specify interfaces for unique asset representation, transfer semantics, and metadata handling~\cite{wang2021non}. Foundational proposals such as ERC-721~\cite{eip721} established the blueprint for unique, indivisible tokens, while ERC-1155~\cite{eip1155} introduced a multi-token interface enabling both fungible and non-fungible assets within a single contract.

\smallskip
\noindent\textbf{Yet, we still have an insufficient level of understanding!} 
Prior studies examined NFT markets from a variety of perspectives. A line of works have focused on \textit{economic valuation}, employing multimodal data to estimate token prices~\cite{costa2023show,niu2024unveiling}. Others have investigated \textit{social influence}, such as the role of Twitter in shaping NFT value perception~\cite{kapoor2022tweetboost}, or the behavioral dynamics in NFT speculation and rug pulls~\cite{huang2023miracle}. Another line of research has explored \textit{security}, identifying smart contract defects~\cite{yang2023definition}, wallet hijacking~\cite{stoger2023demystifying}, and airdrop manipulation using graph-based learning~\cite{zhou2024artemis}. In addition, cross-domain research has sought to investigate broader \textit{sociotechnical} aspects of NFTs, including centralization risks~\cite{xiao2024centralized}, NFT identity and provenance~\cite{chen2024towards}.

We argue that the underlying \textbf{\textit{standard layer}} of NFTs received limited attention. Most existing works treat NFTs as black-box tokens deployed in markets, without dissecting how they are structured, which standards they implement, and how these standards evolve, overlap, or fragment (cf. Table~\ref{tab:related-work}).

\begin{table}[!hbt]
\centering
\caption{This work \textcolor{gray}{v.s.} Prior NFT Studies}\label{tab:related-work}
\resizebox{\linewidth}{!}{
\begin{tabular}{>{\columncolor{gray!10}}c cccc}
\toprule
\textbf{refs} & \textbf{Method} & \textbf{Focus} & \textbf{EIP-Level} & \textit{\textbf{Scale}} \\
\midrule

\cite{costa2023show}\cite{niu2024unveiling} &  ML models & Price prediction & \textcolor{red}{\xmark} & \textcolor{red}{$<$10K tokens} \\

\cite{kapoor2022tweetboost}\cite{huang2023miracle} & Behavioral analysis & Social influence & \textcolor{red}{\xmark} & \textcolor{red}{$<$50K tweets / wallets} \\

\cite{yang2023definition}\cite{zhou2024artemis} & Graph learning & Contract defects & \textcolor{orange}{Partial} & \textcolor{orange}{$\sim$1K--500K contracts/wallets} \\

\cite{stoger2023demystifying}\cite{xiao2024centralized} & Case studies & Wallet risks & \textcolor{red}{\xmark} & \textcolor{red}{Dozens of cases} \\

\cite{chen2024towards} & Qualitative analysis & NFT provenance & \textcolor{red}{\xmark} & \textcolor{red}{Dozens of stakeholders} \\

\midrule

\textit{\textbf{This}} & EIP empirical & \textcolor{blue}{NFT standards} & \textcolor{blue}{\cmark} & \textcolor{blue}{$>$10K \{EIPs + forum\}} \\

\bottomrule
\end{tabular}
}
\end{table}

The oversight hinders a deeper understanding of several \textit{key questions}: What is the true scope of NFT-related EIPs? Which functionalities have become foundational versus rarely adopted? How are inheritance and interface structures organized across proposals? Who drives their development? What socio-technical patterns shape participation? What major vulnerability vectors persist across widely adopted and emerging standards, threatening the robustness of the NFT ecosystem? 

We open the \textit{black box} of NFT standards and empirically mapping their structural and functional landscape.

\smallskip
\noindent\textbf{Our approach.}
We empirically evaluate NFT-related EIPs from multi-dimensions (\textcolor{magenta}{\S\ref{sec-methodology}}).
We first construct a structured dataset by downloading the entire corpus of EIPs and applying keyword-based filtering to identify 213 NFT-related proposals, out of which 191 contain explicit Solidity interface definitions. We extract proposal metadata (stage, creation date, deadlines) and parse interface-level information (functions, parameters, return types) to map the functional design space of NFT standards and their inheritance relationships.  We complement this technical analysis with contributor profiling, linking authors to publicly available GitHub data to characterize geographic and organizational participation patterns. Additionally, we collect community engagement data from \underline{Ethereum Magicians}\footnote{\url{https://ethereum-magicians.org/}, a long-standing open discussion forum where Ethereum community members propose, debate, and refine EIPs} forum threads, capturing temporal dynamics of discussions and participation levels of NFT-related proposals.

To extend our analysis beyond structural, functional, and sociotechnical aspects, we incorporate two complementary investigations (\textcolor{magenta}{\S\ref{sec-extend}}). 
First, we compile a corpus of prominent academic studies on NFTs and cross-reference their scope, methods, and targeted standards against our dataset to assess the alignment between \textit{existing research efforts} and the actual landscape of NFT-related EIPs. Second, we conduct a focused review of fundamental NFT standards (e.g., ERC‑721, ERC‑1155, ERC‑6551) and emerging proposals (e.g., ERC‑4907, ERC‑4675, ERC‑2981) to identify intrinsic \textit{security challenges} and \textit{vulnerability vectors}. 
This combined perspective provides a deeper understanding of how academic work connects to real-world standardization processes.

\smallskip
\noindent\textbf{Contributions.}
We provide a comprehensive and data-driven perspective on the evolution of Ethereum NFT standards. 

\noindent\hangindent 1em\textit{$\triangleright$ A structural landscape of NFT standards} (\textcolor{magenta}{\S\ref{sec-rq1}}-\textcolor{magenta}{\S\ref{sec-rq4}}). 
We provide the first systematic mapping of NFT-related EIPs, characterizing their distribution across proposal stages, their temporal evolution, and the diversity of functionalities they introduce. Our analysis identifies foundational proposals that have shaped the NFT ecosystem.

\smallskip
\noindent\hangindent 1em\textit{$\triangleright$ A socio-technical perspective on NFT standardization} (\textcolor{magenta}{\S\ref{sec-rq5}}). 
We examine the human and organizational dynamics driving NFT standardization, including contributor anonymity, geographic distribution, and community engagement levels in the EIP discussion process. This perspective highlights how institutional and collaborative factors influence the trajectory and adoption of NFT standards beyond purely technical considerations.

\smallskip
\noindent\hangindent 1em\textit{$\triangleright$ A synthesis of academic coverage and security challenges} (\textcolor{magenta}{\S\ref{sec-rq6}}-\textcolor{magenta}{\S\ref{sec-rq7}}). 
We bridge the gap between empirical findings on NFT standards and existing academic literature. We show that prior research primarily focused on markets and selected early standards, leaving later proposals and protocol-level design questions largely unexplored. Furthermore, we identify security challenges intrinsic to major NFT standards and emerging proposals.







\section{Related NFT Studies}
\label{sec-relatedwk}

\subsection{Security Vulnerability and Market Integrity}

A significant body of research has focused on the security and integrity of the NFT ecosystem, spanning from smart contract vulnerabilities to widespread market manipulation. At the contract level, researchers have provided systematic overviews of security issues~\cite{von2022nft, das2022understanding}, developed tools to automatically detect defects in NFT smart contracts~\cite{yang2023definition, ma2025uncovering, he2023unknown}, and targeted specific exploits like \textit{sleepminting} with specialized detection tools~\cite{xiao2025wakemint}. More recent work~\cite{wang2025ai} has applied AI models to identify common NFT vulnerabilities such as risky proxies and reentrancy.

Beyond contract flaws, studies have examined systemic threats to market integrity. This includes large-scale measurement of \textit{NFT rugpulls}~\cite{huang2023miracle}, where developers abandon projects after securing investor funds, and the development of graph learning systems like ARTEMIS to detect manipulative \textit{airdrop hunters}~\cite{zhou2024artemis}. However, the most extensively studied threat is \textit{wash trading}. Foundational work by von Wachter et al.~\cite{von2022nft} quantified this issue, finding that 2.04\% of sales transactions in their sample were suspicious, inflating volume by nearly \$150 million. La Morgia et al.~\cite{la2023game} provided a larger-scale analysis, estimating that 5.66\% of all Ethereum collections were affected by wash trading, creating over \$3.4 billion in artificial volume. A key insight from this research is that the motivation is often not just price inflation but also reward farming on marketplaces that incentivize trading volume; Niu et al.~\cite{niu2024unveiling} found that on such platforms, wash trading could account for as much as 94.5\% of all activity. To combat this, researchers have proposed various detection methods, including graph-based approaches that identify cyclical trading patterns~\cite{tahmasbi2023identifying} and visual analytics tools that empower human experts to spot manipulation schemes~\cite{wen2023nftdisk}.

\subsection{Market Dynamics and Ecosystem Analysis}

Another major research area involves characterizing NFT market dynamics. Early work applied machine learning to predict NFT prices based on visual and textual features~\cite{costa2023show} and demonstrated that social media activity is a significant factor in valuation models~\cite{kapoor2022tweetboost}. To support such analysis, resources like ``Live Graph Lab'' have been created to provide large-scale, dynamic graphs of NFT transactions~\cite{zhang2024live}. 

More recent studies have focused on characterizing specific ecosystems beyond Ethereum. For instance, White et al.~\cite{white2022characterizing} conducted a detailed analysis of the OpenSea marketplace, one of the largest by volume. In parallel, Kong et al.~\cite{kong2024characterizing} performed the first systematic study of the Solana NFT ecosystem, revealing a strong correlation between NFT sales and the price of Solana's native token (SOL). Building on this ecosystem analysis, application-layer research aims to improve user experience. Yu et al.~\cite{yu2023predicting} proposed a recommender system for Web3 assets that uses a Graph Neural Network (GNN) to predict NFT classifications, addressing the challenge of navigating vast and complex collections. Studies ~\cite{xiao2024centralized, stoger2023demystifying, heimbach2023defi} have found that despite the ``trustless'' ethos of Web3, users actively seek centralized trust anchors, and that stakeholder value is derived not just from financial aspects but also from utility, community, and provenance~\cite{chen2024towards, baytacs2022stakeholders}.

\subsection{Emerging Standards}
While much research analyzes existing systems, a forward-looking stream of work focuses on evolving the foundational NFT protocols themselves. A key limitation of the standard ERC-721 token is its isolated nature, which offers creators only one-time incentives. To address this, Wang et al.~\cite{wang2023referable} proposed the Referable NFT (rNFT) scheme, now standardized as EIP-5521. Further, Yu et al.~\cite{yu2025maximizing} explored this standard by using game theory and deep reinforcement learning to demonstrate how creators can maximize long-term rewards through this network of references. 

Innovation is also occurring outside the Ethereum ecosystem, particularly on Bitcoin~\cite{wang2025understanding,gogol2024writing,li2024bitcoin,wang2024bridging}. Research has provided foundational explorations of Bitcoin Inscriptions, which leverage Ordinal Theory to create NFT-like assets on the Bitcoin blockchain~\cite{li2024bitcoin}. This led to the experimental BRC-20 token standard, which Wang \& Yu analyzed, concluding that while it expands Bitcoin's functionality, it may not match the complexity of Ethereum's dApp ecosystem~\cite{wang2025understanding}. The subsequent "inscription boom" served as a real-world stress test for Layer-2 rollups, with Gogol et al. finding that the surge in transactions paradoxically lowered median gas fees on ZK-rollups due to cost amortization~\cite{gogol2024writing}. To address the isolation of this new Bitcoin-native ecosystem, researchers like Wang et al. have proposed lightweight bridges to connect BRC-20 assets to Ethereum, enabling cross-chain interoperability~\cite{wang2024bridging}.

\section{EIP and NFT}
\label{sec-bck}

\subsection{Ethereum Improvement Proposals}

EIPs are formal design documents that define proposed changes to the Ethereum blockchain and its surrounding ecosystem. They constitute the primary coordination mechanism for introducing new features, improving protocol-level behavior, and standardizing application-level interfaces~\cite{wang2021non}. EIPs provide a unified template for presenting specifications, justifying design choices, and enabling community-wide evaluation prior to implementation. EIPs have governed both critical protocol transitions, such as the post-DAO hard fork (EIP-779~\cite{eip779}), and the evolution of smart contract standards that underpin decentralized applications and token economies.

\smallskip
\noindent\textbf{Stages.}
The EIP lifecycle follows a well-defined governance process. The process begins with the \textit{Idea} stage, where contributors formulate potential improvements and share them on community forums (i.e., Ethereum Magicians). Upon submission as a properly formatted draft to the GitHub-based EIP repository, the proposal enters the \textit{Draft} stage and is assigned a unique EIP number. The draft remains mutable and subject to iterative refinement.

Following broader community feedback, a draft may transition to \textit{Review}, initiating a formal commentary period lasting at least 45 days. If consensus emerges and no critical objections remain, the proposal proceeds to the \textit{Last Call} phase, a final checkpoint for minor adjustments. Proposals that successfully complete Last Call may be marked as \textit{Final}, indicating readiness for adoption in clients or standards libraries. EIPs that become obsolete, lack sufficient engagement, or are formally abandoned are categorized as either \textit{Stagnant} (inactive for 6+ months) or \textit{Withdrawn} (voluntarily retracted).

\smallskip
\noindent\textbf{Types.}
EIPs are formally categorized into six distinct types, each reflecting a different layer of the protocol stack.

\textit{Standards track EIPs} are binding proposals that affect either the core protocol or its application-layer interfaces. This umbrella category is further subdivided into four subtypes:

\begin{itemize}
    \item \textbf{Core.} These proposals modify Ethereum’s consensus rules or client state transition logic. They typically require coordinated network upgrades and are consensus-critical (e.g., EIP-1559 on fee market reform).
    
    \item \textbf{Networking.} Focused on peer-to-peer messaging and Ethereum’s devp2p stack, these EIPs affect how clients communicate but do not alter consensus behavior (e.g., EIP-1459 for node discovery).
    
    \item \textbf{Interface.} These define external client-facing APIs, such as JSON-RPC methods or CLI conventions (e.g., EIP-1898 for block parameter queries).
    
    \item \textbf{ERC.} These define standardized application-level smart contract interfaces, including tokens, metadata, royalties, and naming schemes (e.g., ERC-721, ERC-1155).
\end{itemize}

\textit{Meta EIPs}, also referred to as Process EIPs, govern the procedures, tooling, and governance mechanisms behind Ethereum’s protocol evolution. These proposals shape the EIP workflow itself or meta-level policy decisions (e.g., EIP-1 for EIP formatting rules, EIP-5069 for draft status transitions).

\textit{Informational EIPs} serve as non-binding documentation of community best practices, architectural rationales, or historical context. Although not enforced or consensus-driven, these proposals contribute to transparency and institutional memory (e.g., EIP-6953 on upgrade activation schemes).

\begin{table}[t]
\centering
\caption{Ethereum EIP Classification Overview}
\label{tab:eip-types}
\begin{threeparttable}
\resizebox{\linewidth}{!}{%
\begin{tabular}{>{\columncolor{gray!10}}c cccc}
\toprule
\textbf{EIP Type} & \textbf{Binding} & \textbf{Scope} & \textbf{Consensus} & \textbf{Example(s)} \\
\midrule
\textbf{Core}          & \textcolor{blue}{Yes} & Consensus Layer   & \textcolor{green!50!black}{\textbf{Required}} & EIP-1559, EIP-3675 \\
\textbf{Networking}    & \textcolor{blue}{Yes} & P2P Protocols     & \textcolor{green!50!black}{\textbf{Required}} & EIP-1459, EIP-868 \\
\textbf{Interface}     & \textcolor{blue}{Yes} & RPC / APIs        & \textcolor{green!50!black}{\textbf{Required}} & EIP-1898, EIP-1474 \\
\textbf{ERC}           & \textcolor{blue}{Yes} & Smart Contracts   & \textcolor{green!50!black}{\textbf{Required}} & ERC-20, -721, -1155 \\
\textbf{Meta}          & \textcolor{blue}{Yes} & Governance Process& \textcolor{green!50!black}{\textbf{Required}} & EIP-1, EIP-5757 \\
\textbf{Informational} & \textcolor{red}{No}   & Best Practices    & \textcolor{orange!85!black}{\textbf{Optional}} & EIP-6953, EIP-2333 \\
\bottomrule
\end{tabular}%
}
\begin{tablenotes}
\footnotesize
 \item[]   \textbf{\textit{Binding}} indicates whether a proposal must be implemented or followed
 \item[]    after acceptance. \textbf{\textit{Consensus}} refers to the requirement for coordinated
 \item[]    network/client adoption. Subtypes under \textit{Standards track} are consensus-
 \item[]   bound and must be included in client releases once finalized.
\end{tablenotes}
\end{threeparttable}
\end{table}

\smallskip
\noindent\textbf{Exemplified EIPs.}
To illustrate the functional evolution, we examine two canonical EIPs~\cite{wang2021non} that have fundamentally shaped the NFT landscape: \textbf{EIP-721} and \textbf{EIP-1155}.

\textbf{\ding{202} EIP-721}, formally titled \textit{Non-Fungible Token Standard}, introduced in early 2018 and finalized later that year, defines the foundational interface for managing unique, indivisible digital assets. It was the first widely adopted standard to encode token-level uniqueness, enabling verifiable ownership and transfer of discrete digital items across Ethereum dApps.

Technically, ERC-721 establishes a set of required and optional functions that must be implemented by compliant smart contracts. These include:
\begin{itemize}
    \item \texttt{ownerOf(tokenId)}, \texttt{balanceOf(owner)}: functions to track individual ownership and user holdings;
    \item \texttt{transferFrom(from, to, tokenId)} and its safe variant to securely transfer token ownership;
    \item \texttt{approve(to, tokenId)}, \texttt{setApprovalForAll(operator, approved)} for delegation and permissioned asset control;
    \item Interface support via ERC-165 and optional metadata (\texttt{name()}, \texttt{symbol()}, \texttt{tokenURI()}) and enumeration (\texttt{totalSupply()}, \texttt{tokenOfOwnerByIndex()}, etc.).
\end{itemize}

\textbf{\ding{203} EIP-1155} generalizes the notion of token identity by introducing a unified contract interface for managing multiple token types, both fungible and non-fungible, within a single smart contract. This composite architecture significantly reduces contract deployment overhead and allows for highly gas-efficient batch operations, making it particularly suitable for blockchain games and metaverse applications.

The core design pivots around the notion of token type IDs (\texttt{uint256 id}), each associated with a supply count: 

\begin{itemize}
    \item \texttt{balanceOf(account,id)}/\texttt{balanceOfBatch(accounts,ids)}: efficient balance queries across tokens and users;
    \item \texttt{safeBatchTransferFrom()} and \texttt{safeTransferFrom()}: native support for batch transfers and transfer safety;
    \item URI substitution: metadata retrieval via a single endpoint supporting multiple token types (\texttt{uri(id)}), optimized for off-chain storage;
    \item Enhanced safety via \texttt{IERC1155Receiver} interface for secure contract-to-contract interaction.
\end{itemize}

\subsection{NFTs in Practice}
NFTs are cryptographic assets that represent unique, indivisible items on the blockchain. Unlike fungible tokens (e.g., ERC-20) that are interchangeable by quantity, NFTs are used to certify ownership of digital or physical assets, ranging from digital art and collectibles to identity credentials and access rights. Each NFT is associated with a distinct identifier, typically referred to as \texttt{tokenId}, and governed by smart contract standards such as ERC-721 or ERC-1155.

\smallskip
\noindent\textbf{Foundational design.}
NFTs follow a hybrid design that separates on-chain ownership tracking from off-chain media storage.  At the core of an NFT’s representation is a unique identifier, \texttt{tokenId} $\in \mathbb{Z}^+$, which is used to associate a digital asset with its owner and related metadata. On-chain, this association is maintained via mappings, which together constitute the state structure of an ERC-721-compliant smart contract:

\begin{lstlisting}[language=Solidity]
mapping(uint256 => address) private _owners;
mapping(address => uint256) private _balances;
mapping(uint256 => string)  private _tokenURIs;
\end{lstlisting}

Here, \texttt{\_owners} maps each \texttt{tokenId} to its current holder, while \texttt{\_balances} tracks how many NFTs each address owns. \texttt{\_tokenURIs} acts as a pointer (usually an IPFS hash or HTTPS URL) to a structured JSON object that describes the asset’s attributes. While this metadata is not stored directly on-chain, it forms a canonical interface for marketplaces, wallets, and indexers to render NFT contents consistently. A typical metadata object may resemble:

\begin{lstlisting}[language=json]
{
    "name": "CryptoArtifact #42",
    "description": "A unique digital relic.",
    "image": "ipfs://QmX…42.png",
    "attributes": [
        { "trait_type": "Rarity", 
        "value": "Epic" }]
}
\end{lstlisting}

\smallskip
\noindent\textbf{Minting, transfer, and approval.}
The ERC-721 standard specifies a set of primitive operations that govern the lifecycle of a non-fungible token. These operations include minting a new token, authorizing a third party, and executing transfers. 

\begin{lstlisting}[language=Solidity]
// Token ownership and approvals
mapping(uint256 => address) private _owners;
mapping(uint256 => address) private _tokenApprovals;

// Minting: create a new NFT and assign to add_to
function _mint(address to, uint256 tokenId) internal {
    require(_owners[tokenId] == address(0));
    _owners[tokenId] = to;
    emit Transfer(address(0), to, tokenId);
}

// Approve another address to transfer tokenId
function approve(address to, uint256 tokenId) public {
    require(msg.sender == _owners[tokenId]);
    _tokenApprovals[tokenId] = to;
    emit Approval(msg.sender, to, tokenId);
}

// Transfer NFT from add_from to add_to
function transferFrom(address from, address to, uint256 tokenId) public {
    require(msg.sender == _owners[tokenId] || msg.sender == _tokenApprovals[tokenId]);
    _owners[tokenId] = to;
    emit Transfer(from, to, tokenId);
}
\end{lstlisting}

\begin{figure}[!t]
    \centering
        \includegraphics[width=0.99\linewidth]{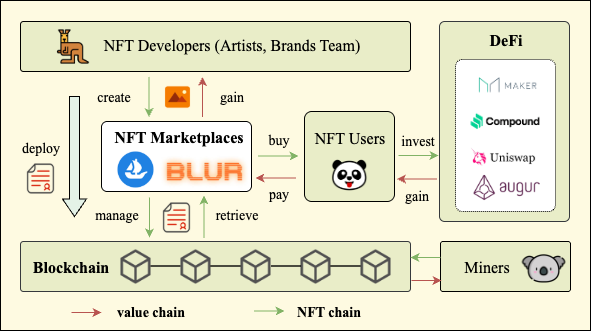}
    \caption{NFT Value Chain Flow}
    \label{fig:value-chain}
\end{figure}

\smallskip
\noindent\textbf{NFT value chain.}
The value chain of NFTs involves a series of stages that reflect how these digital assets are created, exchanged, and utilized (\textcolor{teal}{Fig.~\ref{fig:value-chain}}). 

The first stage is \textit{creation}, minted by artists, developers, or brands. This involves deploying a smart contract that generates a new token ID and links it to metadata. Once minted, NFTs enter the \textit{circulation} stage. They are typically listed on public marketplaces (e.g., OpenSea~\footnote{\url{https://opensea.io/}, one of the largest NFT marketplaces by trading volume} and Blur~\footnote{\url{https://blur.io/}, a professional NFT trading platform known for fast bulk listings and bidding features}), where they can be sold through fixed prices or auction mechanisms. The transfer of ownership is handled by on-chain smart contracts. 

After a successful transaction, NFTs reach the \textit{consumption} stage. Holders may use them in many ways. For example, gaining access to games, exclusive content, or online communities. NFTs can also represent social identity or serve as speculative investments, with owners hoping to profit from future resale. 

Finally, NFTs often undergo \textit{secondary transfer} or \textit{financialization}. This includes resale on secondary markets, as well as emerging use cases such as collateralized lending, NFT staking, and fractional ownership. These mechanisms allow NFTs to expand their role beyond simple collectibles.

\section{Methodology}
\label{sec-methodology}

\subsection{Research Questions (RQs)}
To provide a data-driven understanding of NFT standardization, we identify a set of research questions that span multiple analytical levels, ranging from macro-level standard evolution to micro-level function usage and parameter structures:

\smallskip
\noindent\hangindent 1em\textbf{$\triangleright$ RQ1: What is the overall landscape of NFT-related EIPs?}  
This question seeks to characterize the general scope and development trajectory of NFT standards. Specifically, we investigate: (1) the total number of NFT-related EIPs proposed to date; (2) their distribution across different proposal stages (e.g., \textit{Draft}, \textit{Review}, \textit{Final}, \textit{Stagnant}); and (3) the temporal evolution of NFT EIPs, measured by the number of proposals introduced over time.

\smallskip
\noindent\hangindent 1em\textbf{$\triangleright$ RQ2: How are functionalities defined and adopted across NFT standards?}  
We examine the diversity of function-level definitions across EIPs. This includes: (1) identifying the total number of distinct NFT-related functions defined in the standard corpus; (2) analyzing the distribution of these functions by category; and (3) evaluating the frequency of function reuse across EIPs to determine if certain functions act as cornerstones within the ecosystem.

\smallskip
\noindent\hangindent 1em\textbf{$\triangleright$ RQ3: What inheritance relationships exist among NFT standards?}  
To investigate the structural organization of NFT-related EIPs, we construct an inheritance graph in which nodes represent individual standards and directed edges capture explicit inheritance and extension relationships declared within the EIP corpus. 

\smallskip
\noindent\hangindent 1em\textbf{$\triangleright$ RQ4: What are the parameter-level characteristics of NFT standards?}  
This question addresses how NFT-related EIPs define their functional interfaces, with a focus on the types, composition, and co-occurrence of input and output parameters. By analyzing function signatures across all standards, we quantify the distribution of recurring primitives in different function classes and uncover dominant pairing patterns among parameters.

\smallskip
\noindent\hangindent 1em\textbf{$\triangleright$ RQ5: What are the socio–geographic and participation patterns of NFT EIP contributors?}
To characterize how human contributors shape the NFT standardization process, we analyze contributor–standard relations along several dimensions: (1) the prevalence of anonymity or pseudonymity among authors; (2) the geographic distribution of identified contributors, summarized by countries/regions; (3) the distribution of standards by the number of authors per EIP; (4) the distribution of authors by the number of standards to which they contribute; and (5) the role of organizations.

\smallskip
\noindent\hangindent 1em\textbf{$\triangleright$ RQ6: How do existing academic studies align with the NFT standard landscape?}  
To contextualize our findings within prior literature, we survey prominent research works that focus on NFTs and examine: (1) how many distinct research threads have emerged in existing studies; and (2) which NFT standards are predominantly targeted or referenced within these lines of research. 

\smallskip
\noindent\hangindent 1em\textbf{$\triangleright$ RQ7: What are the key security challenges and vulnerability vectors across major NFT standards?}
As an extension of our analysis, we examine prominent NFT-related EIPs (ERC‑721, ERC‑1155, ERC‑6551), and emerging proposals like ERC‑4907, ERC‑4675, and ERC‑2981 to identify their intrinsic security weaknesses, and examine how evolving EIP designs introduce new attack surfaces.

\subsection{Data Collection Process}
To support our empirical study of NFT standardization, we developed a comprehensive dataset that integrates Ethereum official documentation, Solidity interface structures, contributor metadata, and community discussion records. 

\smallskip
\noindent\textbf{Standard identification and metadata extraction.} 
The data collection process follows a reproducible and modular pipeline that proceeds in multiple coordinated steps. We first obtained the full corpus of EIPs by downloading all HTML documents from the official repository~\footnote{\url{https://eips.ethereum.org/all}, the canonical EIPs website, hosting authoritative specifications of all proposed and finalized EIPs}. Each proposal was parsed locally to allow for flexible analysis and data extraction. 

To identify NFT-related EIPs, we applied keyword-based filtering using terms including but not limited to “nft” and “non-fungible”. We also applied a second round of investigation manually to confirm their validity. The filtering process yielded a total of 213 EIPs potentially related to NFTs, out of which 191 contained concrete Solidity interface definitions. 

For each selected EIP, we extracted metadata including its current stage (e.g., \textit{Draft}, \textit{Review}, \textit{Final}, \textit{Stagnant}), creation date, and last call deadline, enabling a temporal and categorical analysis of NFT proposal evolution.

\smallskip
\noindent\textbf{Interface parsing and content profiling.}
Building on this metadata, we parsed the embedded Solidity code of each standard using the BeautifulSoup HTML parser and regular expressions, extracting interface declarations, function signatures, parameter types, and return values. This allowed us to construct a structured repository of NFT functionalities and their associated interface design patterns. Additionally, we analyzed the inheritance relationships among interfaces to support dependency graph construction. 

To further characterize contents (e.g., human contributors behind standards), we extracted author information directly from the EIP documents, including names, email addresses, and GitHub usernames. We further enriched this information by querying the GitHub API to retrieve author profiles, company affiliations, and locations, enabling a sociotechnical mapping between standards and contributor networks.

\smallskip
\noindent\textbf{Community discussion mining.}
Beyond technical specifications, we examined community participation surrounding NFT standards. We crawled Ethereum Magicians, focusing on two subforums: \texttt{/c/eips/5} and \texttt{/c/ercs/57}, which host discussions on EIPs and ERCs respectively.  

Using the forum’s public API, we retrieved all discussion topics and associated metadata from 39 pages of EIP threads and 11 pages of ERC threads, yielding a raw dataset of over 10{,}000 discussion entries. NFT-related discussions were identified through pattern matching and cross-referenced with known NFT EIP identifiers. For each matched discussion thread, we collected temporal engagement metrics, including the number of replies, views, unique participants, and timestamps of first and latest activity. 

  \smallskip
  \begin{center}
    \fbox{%
    \begin{minipage}{0.9\linewidth}
    \textbf{Ethical considerations.} We adhered to ethical standards and privacy regulations throughout the data collection process. All data was sourced from publicly available platforms. No personal identifiers were disclosed, and all author-level analyses were aggregated at the institutional or geographic level. This study complies with the EU GDPR and other applicable frameworks.
    \end{minipage}
   }
   \vspace{5pt}
  \end{center}

\subsection{Yielded Datasets}

The outputs of this multi-stage pipeline were exported into structured CSV and XLSM files\footnote{All datasets will be released once this work is finalized.}, each serving a distinct role in our analysis. Table~\ref{tab:nft-datasets} summarizes the major datasets constructed and their corresponding analytical purposes.

\begin{table}[!htb]
\centering
\caption{Overview of NFT Standards Datasets}
\label{tab:nft-datasets}
\scriptsize
\renewcommand{\arraystretch}{0.8}
\setlength{\tabcolsep}{3pt}
\begin{tabularx}{\columnwidth}{@{}%
  >{\columncolor{gray!10}\raggedright\arraybackslash}p{0.32\columnwidth}
  Y                                                          
  >{\centering\arraybackslash}p{0.12\columnwidth}            
  Y@{}}                                                      
\toprule
\multicolumn{1}{c}{\textbf{Dataset}} & \multicolumn{1}{c}{\textbf{Key attributes}} & \textbf{Count} & \multicolumn{1}{c}{\textbf{Purpose}} \\
\midrule
\url{NFT_with_stages.csv} &
\texttt{stage}, \texttt{created\_at}, \texttt{deadline} &
213 &
Standard status tracking and lifecycle analysis \\
\midrule
\url{NFT_interfaces_func.csv} &
\texttt{interface}, \texttt{function}, \texttt{params}, \texttt{returns} &
1572 &
Function structure and inheritance analysis \\
\midrule
\url{NFT_author_infos.csv} &
\texttt{author}, \texttt{github\_profile}, \texttt{affiliation} &
568 &
Contributor attribution and affiliation mapping \\
\midrule
\url{NFT_EIP_topics.xlsm} &
\texttt{views}, \texttt{participants}, \texttt{timestamps}, \texttt{replies} &
121 &
Forum-based community engagement evaluation \\
\midrule
\url{authors_organization.csv} &
\texttt{standard\_symbol}, \texttt{author\_name}, \texttt{email}, \texttt{organization} &
564 &
Mapping contributors to organizations \\
\bottomrule
\end{tabularx}
\vspace{-0.3em}
\end{table}


\section{Empirical Analysis}
\label{sec-emp}
We conduct our investigation on the collected NFT-related EIPs, addressing the research questions \textbf{(RQ1-RQ5)} outlined in \textcolor{magenta}{\S\ref{sec-methodology}}. Table~\ref{tab:result-guidance} summarizes the mapping between each RQ and the figures reporting our main findings.

\begin{table}[!htb]
\centering
\caption{(Result guidance) Mapping RQs to Figures}
\label{tab:result-guidance}
\scriptsize                                   
\renewcommand{\arraystretch}{0.95}          
\begin{tabular}{c|c|l@{}}
\toprule
\multicolumn{1}{c}{\textbf{RQ}} & \multicolumn{1}{c}{\textbf{Index}} & \multicolumn{1}{c}{\textbf{Brief description}} \\
\midrule

\multirow{3}{*}{\textbf{RQ1}} 
& \textcolor{teal}{Fig.~\ref{fig:eip-temporal-trends}(a)} & Monthly NFT-EIPs with 6-month rolling average trend \\
& \textcolor{teal}{Fig.~\ref{fig:eip-temporal-trends}(b)} & Cumulative growth of NFT-EIPs\\
& \textcolor{teal}{Fig.~\ref{fig:stage-distribution}} & Current distribution of NFT-EIPs stages\\
\midrule

\multirow{2}{*}{\textbf{RQ2}} 
& \textcolor{teal}{Fig.~\ref{fig:funCount}} & Categorization of NFT-standard functions \\
& \textcolor{teal}{Fig.~\ref{fig:5_functionfrequency}} & Frequency ranking of functions defined in NFT-EIPs\\
\midrule

\multirow{1}{*}{\textbf{RQ3}} 
& \textcolor{teal}{Fig.~\ref{fig:6_inheritance}} & Temporal evolution of inheritance links \\
\midrule

\multirow{5}{*}{\textbf{RQ4}} 
& \textcolor{teal}{Fig.~\ref{fig:input_types}} & Top 10 most common input parameter types used\\
& \textcolor{teal}{Fig.~\ref{fig:output_types}} & Top 10 return types appearing in NFT functions \\
& \textcolor{teal}{Fig.~\ref{fig:input_output}} & Statistical patterns of input–return parameter combinations \\
& \textcolor{teal}{Fig.~\ref{fig:input_pairs}} & Co-occurrence network of input parameters\\
& \textcolor{teal}{Fig.~\ref{fig:function_param_sankey}} & Sankey diagram linking function names to parameter tags \\
\midrule

\multirow{7}{*}{\textbf{RQ5}} 
& \textcolor{teal}{Fig.~\ref{fig:anonymity}} & Distribution of anonymity levels among NFT-EIPs authors \\
& \textcolor{teal}{Fig.~\ref{fig:bubble_map}} & Geographic distribution of NFT-EIPs authors worldwide\\
& \textcolor{teal}{Fig.~\ref{fig:author_distributions}} & Collaboration network and contribution skew among authors \\
& \textcolor{teal}{Fig.~\ref{fig:nft_eip_orgs}(a)} & Top contributing organizations by the number of NFT-EIPs \\
& \textcolor{teal}{Fig.~\ref{fig:nft_eip_orgs}(b)} & Organizational ranking based on unique contributing authors \\
& \textcolor{teal}{Fig.~\ref{fig:nft_eip_orgs}(c)} & Cumulative NFT-EIPs by leading organizations \\
& \textcolor{teal}{Fig.~\ref{fig:nft_eip_orgs}(d)} & Temporal evolution of NFT-EIPs contributions by organization\\
\bottomrule
\end{tabular}
\end{table}

\subsection{RQ1: The Landscape and Evolution}
\label{sec-rq1}

To understand the development landscape of NFT-related standards on Ethereum, we analyze both the temporal dynamics and the status distribution of 213 NFT-related EIPs. 

\begin{figure}[!htbp]
    \centering
     \subfigure[Monthly number of NFT-related EIPs and 6-month rolling average (2017–2025).]{\label{fig:monthly-count}
     \includegraphics[width=0.9\linewidth]{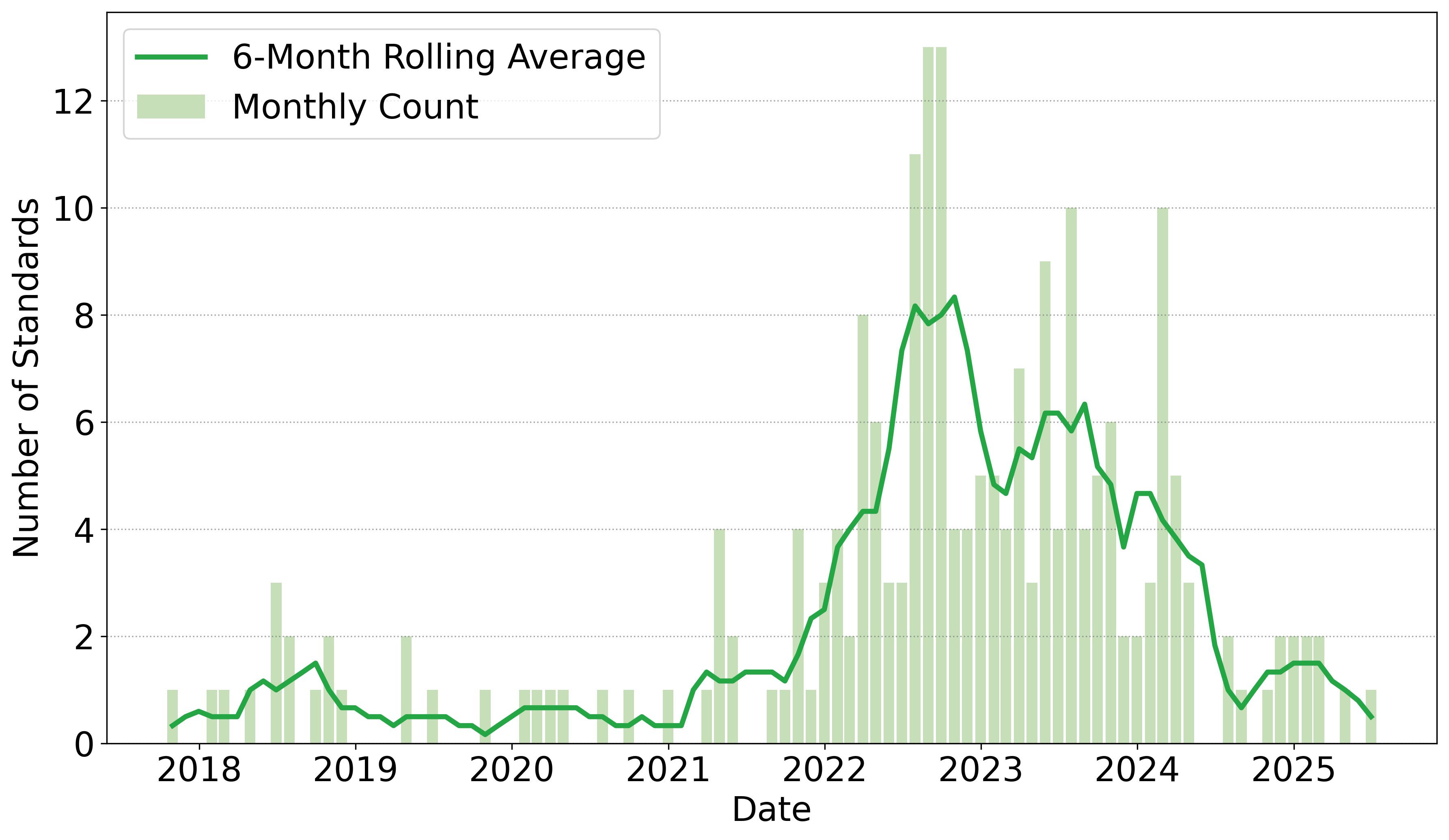}
    }
    \hspace{0.1cm} 
    \subfigure[Cumulative growth of NFT-related EIPs, stratified by proposal stage.]{\label{fig:cumulative-growth}
        \includegraphics[width=0.9\linewidth]{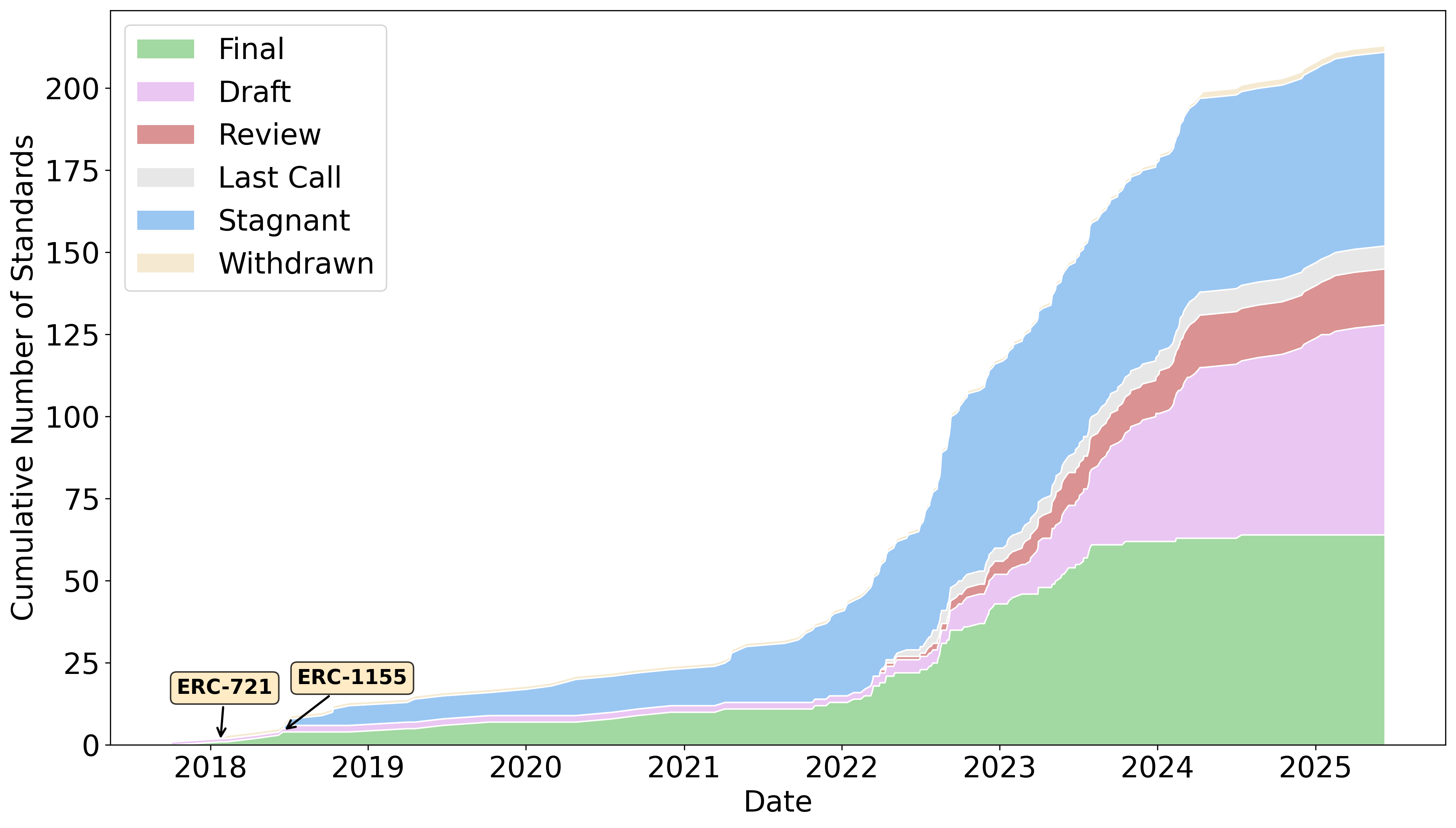}
        }
    \caption{Temporal Trends in the Evolution of NFT-related EIPs.}
    \label{fig:eip-temporal-trends}
\end{figure}

\smallskip
\noindent\textbf{Temporal dynamics of NFT standardization.}
We begin by examining the time series of EIP creation. \textcolor{teal}{Fig.~\ref{fig:monthly-count}} shows the monthly number of NFT-related proposals alongside a 6-month rolling average. In the early period (2017–2019), the rate of NFT-related EIP creation was extremely low, rarely exceeding one or two proposals per month. This phase coincides with the publication of foundational standards such as ERC-721 and ERC-1155, which defined the core abstraction for NFTs but did not immediately spark further standardization. 





A dramatic shift occurred between 2020 and 2022. The monthly creation rate of NFT standards increased sharply, peaking in mid-2022 when more than ten proposals were submitted in some months. The 6-month rolling average confirms sustained high-frequency activity during this period, signaling an ecosystem-wide response to the exponential growth of NFTs in public markets. The rise in proposals correlates with the emergence of NFT marketplaces, mainstream media attention, and the proliferation of NFT-based use cases in digital art, gaming, and DeFi. Correspondingly, \textcolor{teal}{Fig.~\ref{fig:cumulative-growth}} shows a rapid upward trajectory in the total number of NFT-related standards over time, broken down by their stage.

From late 2022 onwards, the standardization pace began to slow. Both the rolling average and monthly counts show a declining trend, and the cumulative growth curve flattens. This reflects a shift in focus from defining new primitives to refining and selectively advancing proposals already in circulation.

\begin{figure}[htbp]
    \centering
    \includegraphics[width=0.75\linewidth]{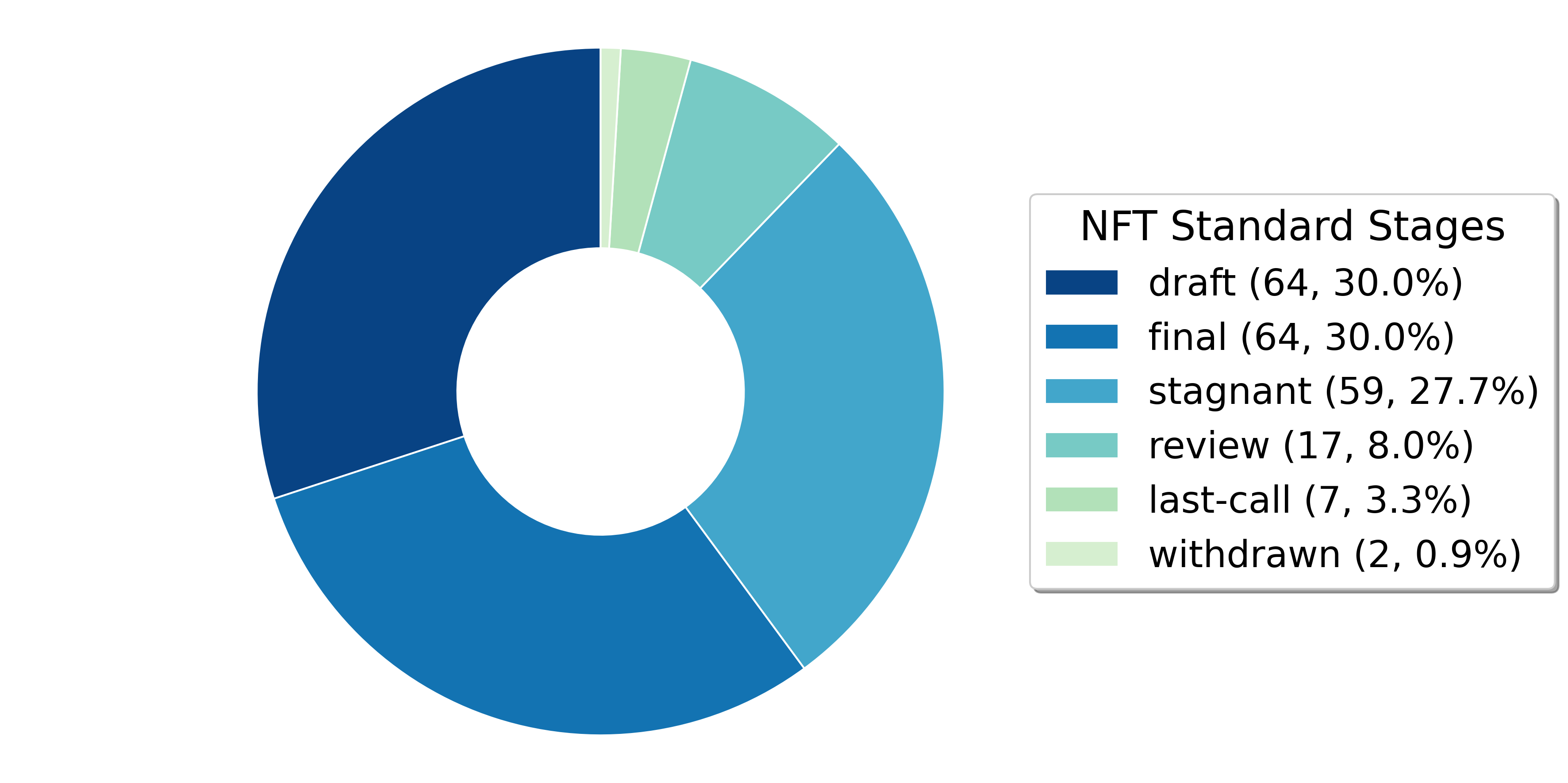}
    \caption{Current distribution of NFT standards by stage.}
    \label{fig:stage-distribution}
\vspace{-0.1in}
\end{figure}

\smallskip
\noindent\textbf{Status distribution.} We examine the current status of all NFT-related EIPs. \textcolor{teal}{Fig.~\ref{fig:stage-distribution}} presents a donut chart showing the proportional distribution across six stages: \textit{Draft}, \textit{Final}, \textit{Stagnant}, \textit{Review}, \textit{Last Call}, and \textit{Withdrawn}.

Two stages dominate the distribution: \textit{Draft} and \textit{Final}, each accounting for 30.0\% of all proposals. This indicates that the NFT standard ecosystem is bifurcated—one half represents fully adopted and mature proposals, while the other half reflects ongoing innovation. Finalized standards, such as ERC-721 and ERC-1155, have become foundational components in the NFT stack, widely supported by tooling and platforms. The large number of draft proposals highlights continued exploration into new NFT functionalities, including fractional ownership, access control, and dynamic metadata.

A substantial portion of proposals (27.7\%) are labeled as \textit{Stagnant}, meaning they have failed to progress. This attrition rate is not uncommon in decentralized ecosystems, where community interest and technical feasibility needs determine the survival of proposals. Nonetheless, the size of this category indicates that while many ideas are proposed, only a subset matures into widely accepted standards.

The remaining proposals fall into intermediate stages: \textit{Review} (8.0\%), \textit{Last Call} (3.3\%), and \textit{Withdrawn} (0.9\%). These imply that only a few proposals are under active consideration. The low percentage of withdrawn proposals suggests that most ideas are either seriously considered or allowed to stagnate naturally, rather than being explicitly rejected.

\begin{center}
\tcbset{
    enhanced,
    boxrule=0pt,
    fonttitle=\bfseries
}
\begin{tcolorbox}[
    lifted shadow={1mm}{-2mm}{3mm}{0.1mm}{black!50!white}
]
\textbf{\text{Insight-\ding{202}:}} The high proportion of draft and stagnant NFT standards reflects an active yet selective standardization pipeline, where a large number of proposals are explored but only a few survive community scrutiny and progress to final adoption.
\end{tcolorbox}
\end{center}

\subsection{RQ2: Functional Definition and Adoption}
\label{sec-rq2}

To investigate how NFT functionalities are designed and reused across standards, we examine three aspects: (1) the categories and distribution of function types; (2) the frequency and popularity of individual functions; and (3) the design conventions in function signatures. 

\begin{figure}[t]
    \centering
    \subfigure[Distribution of NFT–standard functions by category]{
        \begin{minipage}[t]{0.23\textwidth}
            \centering
            \includegraphics[width=\linewidth]{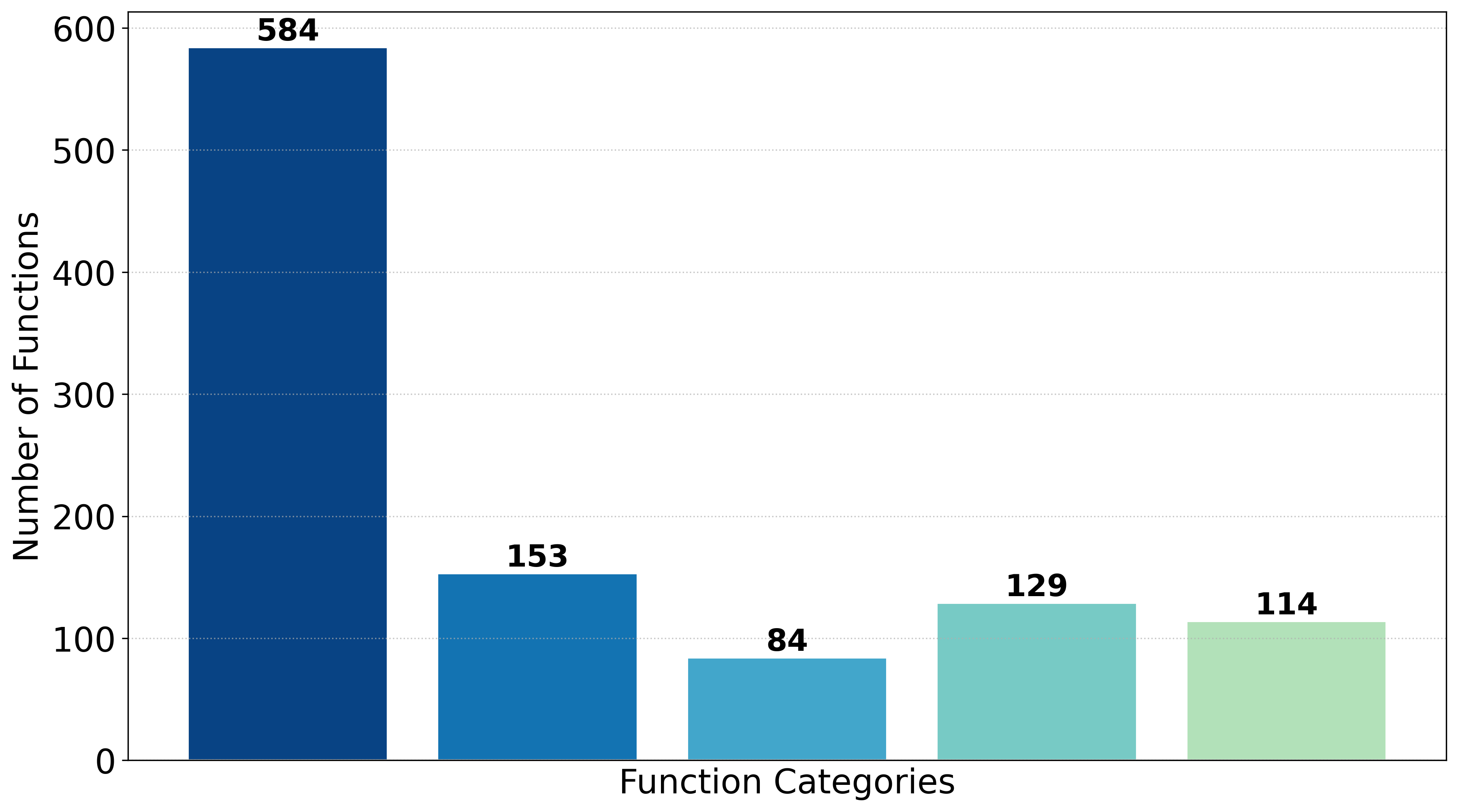}
            \label{fig:funCount}
        \end{minipage}}
    \subfigure[Top functions appearing $>$5 times in ERC proposals]{
        \begin{minipage}[t]{0.23\textwidth}
            \centering
            \includegraphics[width=\linewidth]{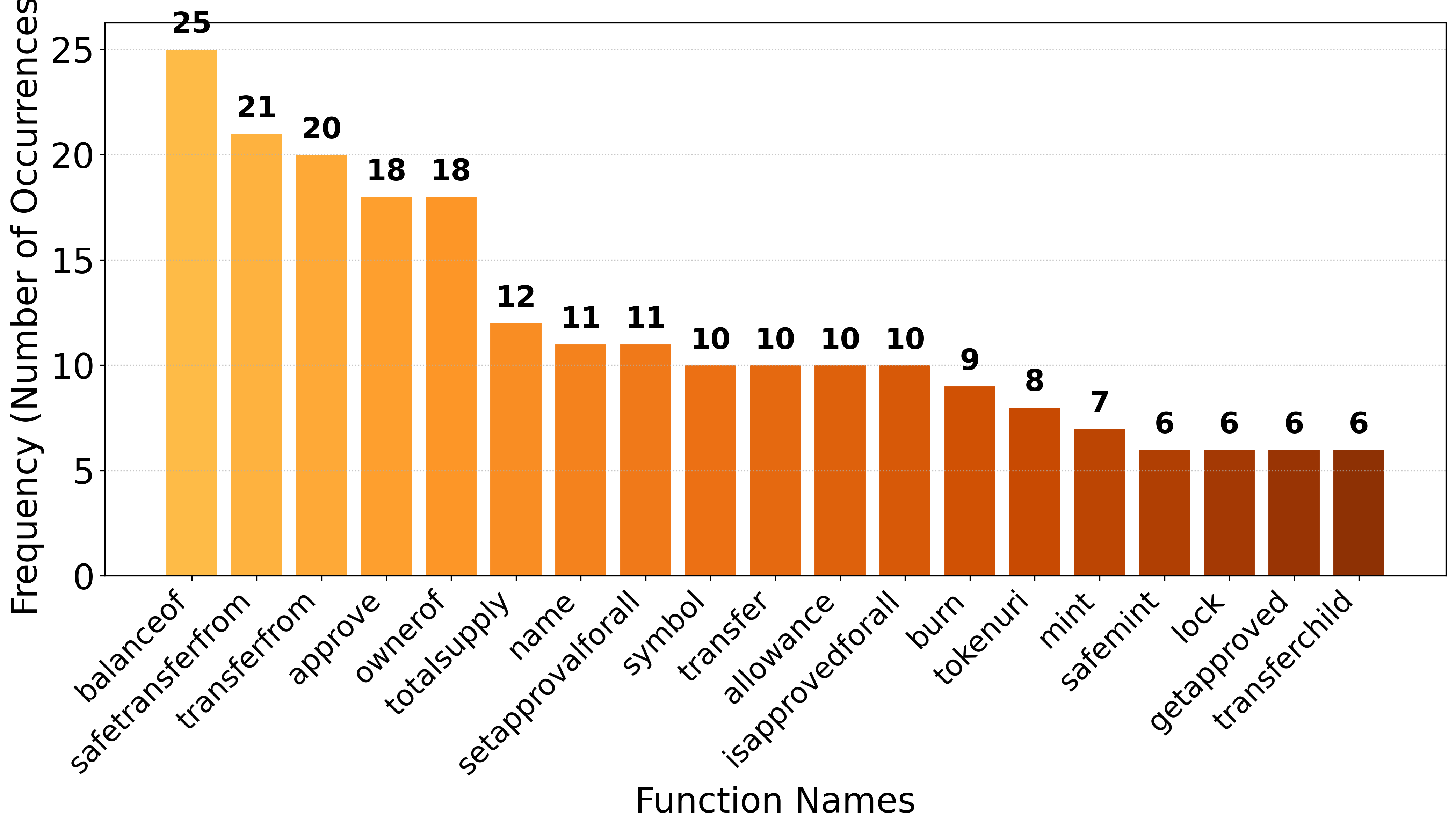}
            \label{fig:5_functionfrequency}
        \end{minipage}}
    \caption{%
        \textbf{Cross-section of function usage in NFT-related ERC interfaces}:
        data management~(\textcolor[rgb]{0.03,0.26,0.52}{\rule{8pt}{8pt}}), %
        relationship\&listing~(\textcolor[rgb]{0.07,0.45,0.70}{\rule{8pt}{8pt}}), %
        registration\&licensing~(\textcolor[rgb]{0.26,0.65,0.80}{\rule{8pt}{8pt}}), %
        access\&permission~(\textcolor[rgb]{0.47,0.79,0.77}{\rule{8pt}{8pt}}), %
        attribute settings~(\textcolor[rgb]{0.70,0.88,0.73}{\rule{8pt}{8pt}}).}
    \label{fig:merged_function_panels}
\end{figure}

\smallskip
\noindent\textbf{Functional categories.} We categorize all defined functions into five major groups based on their intended purpose: \textit{data management}, \textit{relationship and listing}, \textit{access\&permission}, \textit{attribute settings}, and \textit{registration\&licensing}. \textcolor{teal}{Fig.~\ref{fig:funCount}} illustrates the number of unique functions falling under each group.


\smallskip
\noindent\textit{(1) Data management (584 functions).}
This category overwhelmingly dominates the function space, comprising more than half of all NFT-related operations. Data management functions are essential to the fundamental semantics of NFTs, which are inherently data-centric digital assets. These functions typically include mechanisms for querying token balances (e.g., \texttt{balanceOf}), verifying ownership (e.g., \texttt{ownerOf}), retrieving metadata (e.g., \texttt{tokenURI}), and accessing enumeration structures (e.g., \texttt{totalSupply}). Their prevalence reflects the indispensable need for transparency and traceability in NFT applications, such as marketplaces and wallets.

\smallskip
\noindent\textit{(2) Relationship and listing management (153 functions).}
This category encompasses functions that facilitate the association, bundling, or hierarchical organization of NFTs. Examples include operations supporting composability (e.g., \texttt{transferChild}) or group-level actions (e.g., batch listings and transfers). Such mechanisms are particularly useful in domains like gaming, metaverse platforms, and digital asset curation, where individual NFTs often need to be nested within broader structures. The moderate size of this category suggests that while relational complexity is important for certain verticals, it is not universally required across all NFT use-cases.

\smallskip
\noindent\textit{(3) Access and permission controls (129 functions).}
Access control is a critical requirement in any tokenized system, and this category includes functions for defining who can perform which operations. Common examples include \texttt{approve}, \texttt{setApprovalForAll}, and role-based permissioning (e.g., \texttt{isApprovedForAll}). These functions serve to enforce security boundaries within smart contracts, preventing unauthorized modifications or transfers of NFT assets. Their distribution highlights the importance of delegable rights and shared ownership in NFT platforms, especially when assets are managed via custodial wallets and DAOs.

\smallskip
\noindent\textit{(4) Attribute settings (114 functions).}
This group captures functions that define or update specific descriptive properties of an NFT, such as its \texttt{name}, \texttt{symbol}, or metadata URI. Unlike the data management category that focuses on ownership or state queries, attribute-setting functions often relate to presentation and user interface integration. Their moderate presence in the standards corpus indicates that while visual and semantic attributes are important, they are typically defined during initialization and not frequently updated thereafter. 

\smallskip
\noindent\textit{(5) Registration and licensing (84 functions).}
The smallest but still meaningful category, registration and licensing functions are associated with formalizing legal or organizational attributes of NFTs. These may include interfaces for verifying creators, asserting licensing terms, or associating tokens with off-chain registries. Although such functions are not prevalent, their existence reflects a growing recognition of the need for compliance and rights management in enterprise and institutional NFT use-cases. As legal frameworks for NFTs mature, this category may see increased adoption.


\smallskip
\noindent\textbf{Function reuse and popularity.} We analyzed the frequency of function names appearing more than five times in the dataset. As illustrated in \textcolor{teal}{Fig.~\ref{fig:5_functionfrequency}}, the most frequently reused functions reflect the canonical operations expected of NFT contracts.

The most commonly encountered function is \texttt{balanceOf} (25 times), querying the number of tokens owned by a given address and serves as a foundational component for wallets. Transfer operations, including \texttt{safeTransferFrom} (21) and \texttt{transferFrom} (20), also exhibit high frequency. It is not suprising as those functions are the foundation of token transfer. Closely tied to transferability are functions governing ownership and approval, such as \texttt{approve}, \texttt{ownerOf}, \texttt{setApprovalForAll}, and \texttt{isApprovedForAll}. They are used to facilitate permission delegation and third-party access control.

Metadata-related functions, including \texttt{name}, \texttt{symbol}, and \texttt{tokenURI}, appear frequently as well, reflecting the ecosystem’s emphasis on standardized representation of token metadata accessibility. Additionally, supply management functions such as \texttt{totalSupply}, \texttt{mint}, and \texttt{burn} appear with notable regularity, supporting dynamic issuance and lifecycle control of tokens.

Beyond foundational components, our results also reveal the reuse of specialized functions, such as \texttt{lock}, \texttt{safeMint}, \texttt{getApproved}, and \texttt{transferChild}, which point to emerging application domains including composable NFTs, secure minting workflows, and asset-level locking mechanisms. 

\begin{center}
\tcbset{
    enhanced,
    boxrule=0pt,
    fonttitle=\bfseries
}
\begin{tcolorbox}[
    lifted shadow={1mm}{-2mm}{3mm}{0.1mm}{black!50!white}
]
\textbf{\text{Insight-\ding{203}:}} NFT standards exhibit strong convergence around a core set of reusable functions, such as \texttt{balanceOf}, \texttt{transferFrom}, and \texttt{ownerOf}, which serve as foundational primitives across diverse use-cases.
\end{tcolorbox}
\end{center}

\subsection{RQ3: Inheritance Structure among Standards}
\label{sec-rq3}

\begin{figure}[!htbp]
    \centering
    \includegraphics[width=\linewidth]{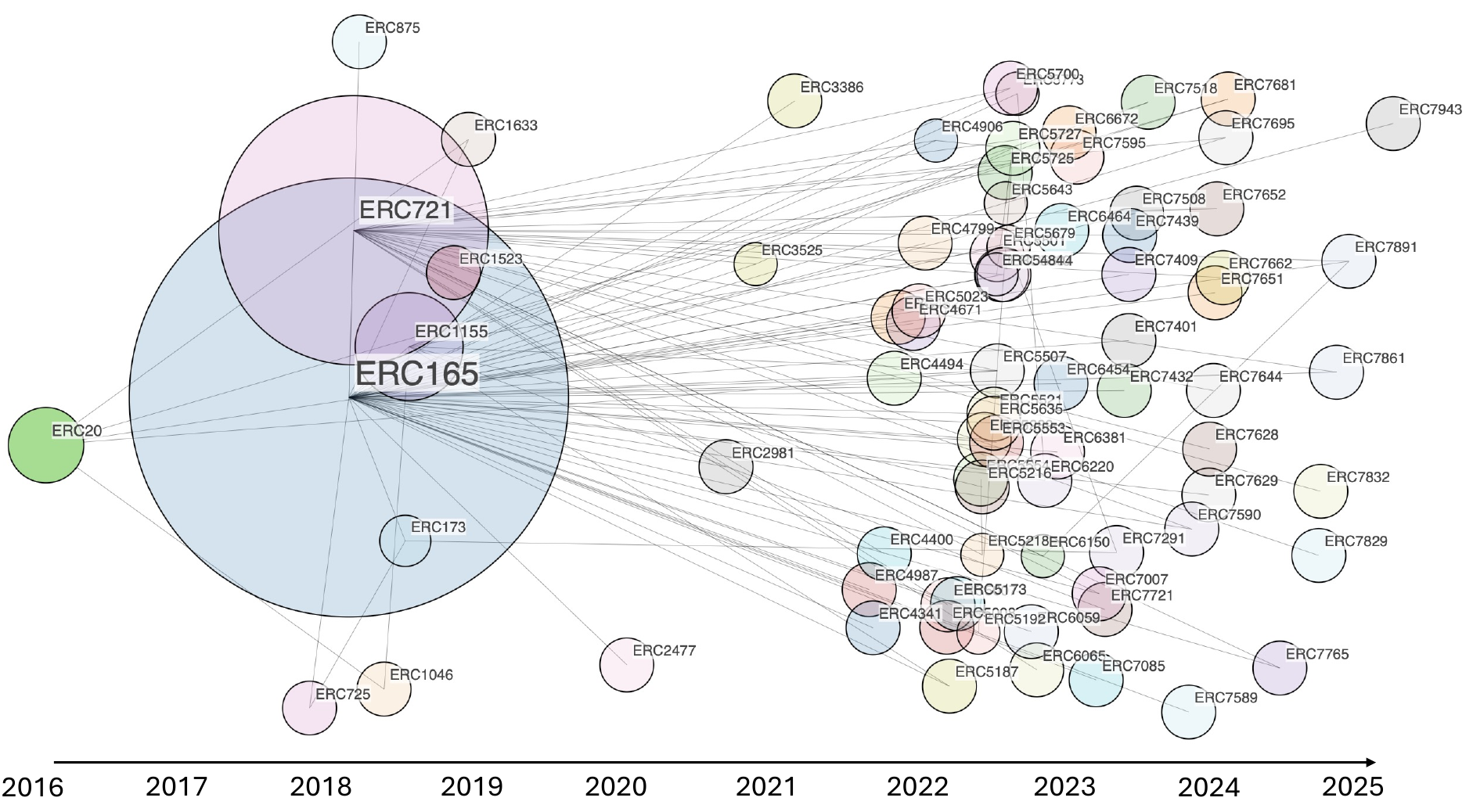}
    \caption{Temporal evolution of EIP inheritance}
    \label{fig:6_inheritance}
\end{figure}

To gain a longitudinal perspective on the evolution of NFT standards, we constructed a time-anchored inheritance graph (\textcolor{teal}{Fig.~\ref{fig:6_inheritance}}), where each node represents an EIP and is positioned along a horizontal axis according to its proposal year. Directed edges capture declared inheritance relationships, and node size encodes in-degree centrality, i.e., the number of standards that explicitly extend or reference it. 


\smallskip
\noindent\textbf{Foundational standards: ERC-165 and ERC-721.}
This visualization reveals several key patterns in the historical development of NFT-related Ethereum standards. First, foundational standards such as \textbf{ERC-165} and \textbf{ERC-721} appear early in the timeline and are represented as large, densely connected hubs. \textit{ERC-165}, introduced in 2018, plays a pivotal role in enabling interface detection and interoperability and is the most widely inherited standard across the graph. Likewise, \textit{ERC-721} serves as the architectural backbone for NFT implementation, forming the basis of numerous later extensions. 

\smallskip
\noindent\textbf{Niche and specialized extensions (2021–2024).}
The period between 2021 and 2024 shows a sharp increase in the number and diversity of NFT standards. While the earlier landscape was dominated by a few high-centrality proposals, the recent years are characterized by a proliferation of smaller, more targeted EIPs, such as \textit{ERC-2981} (royalty specification), \textit{ERC-4907} (rental NFTs), and \textit{ERC-5007} (fractionalized ownership). These standards typically extend from ERC-721 or ERC-1155 and contribute specialized capabilities tailored to specific market demands, such as licensing and DeFi integration.



\begin{center}
\tcbset{
    enhanced,
    boxrule=0pt,
    fonttitle=\bfseries
}
\begin{tcolorbox}[
    lifted shadow={1mm}{-2mm}{3mm}{0.1mm}{black!50!white}
]
\textbf{\text{Insight-\ding{204}:}} The inheritance topology of NFT standards exhibits a core-centric structure, where foundational ERCs like \textit{ERC-165} and \textit{ERC-721} serve as heavily reused architectural primitives. The surge in post-2021 standards signals a phase of specialization and modularization, with new proposals layering targeted functionalities atop well-established interfaces.
\end{tcolorbox}
\end{center}

\subsection{RQ4: Parameter-Level Characteristics of NFT Standards}
\label{sec-rq4}

To investigate the functional primitives underpinning NFT standards, we analyze the input and output parameter types and numbers across all interface-level functions. 







\begin{figure*}[!htbp]
\centering

\subfigure[Top 10 Input Parameter Types]{
\label{fig:input_types}
\includegraphics[width=0.31\linewidth,trim=0 0.8cm 0 0.2cm,clip]{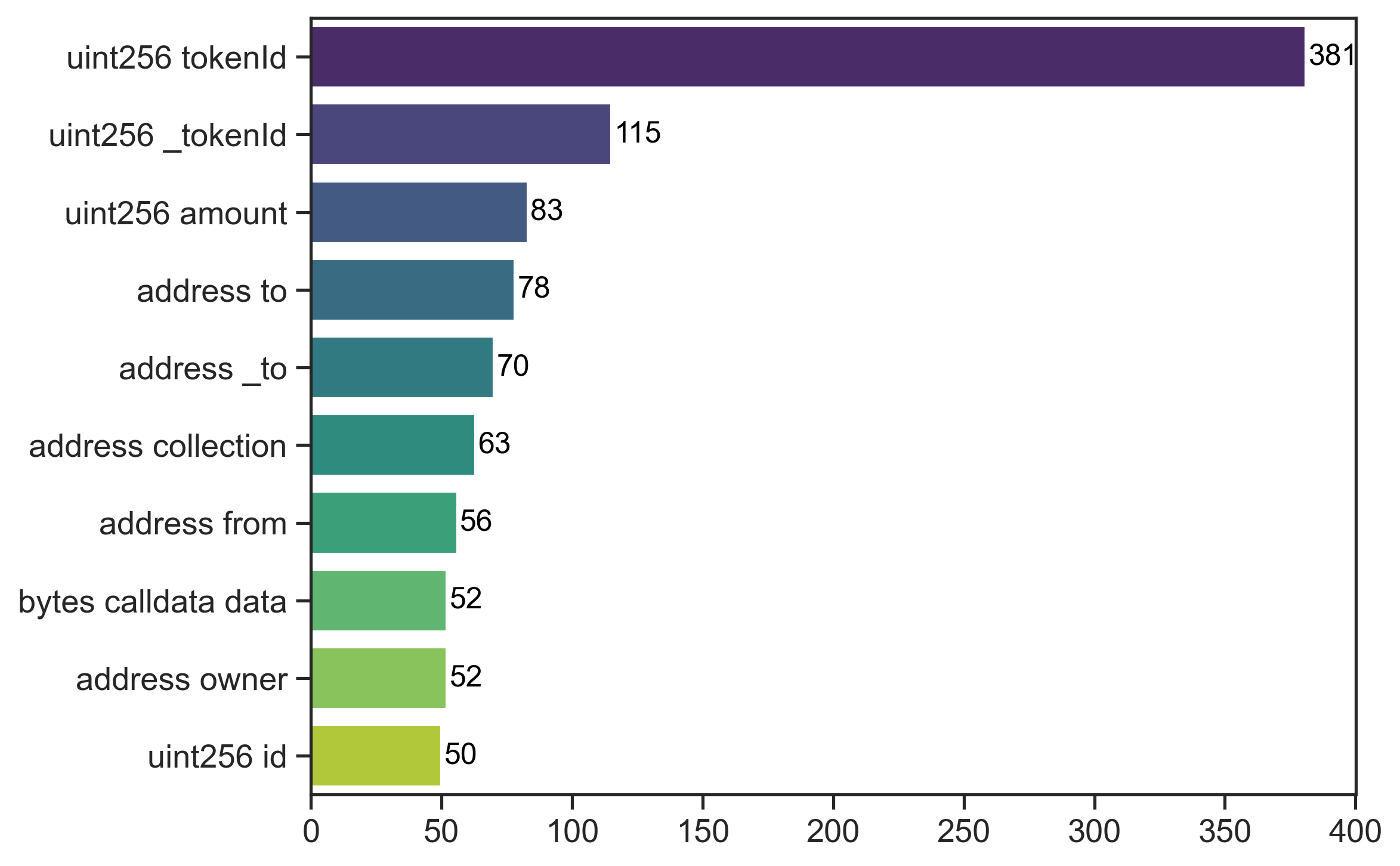}
}
\hfill
\subfigure[Top 10 Return Parameter Types]{
\label{fig:output_types}
\includegraphics[width=0.31\linewidth,trim=0 0.8cm 0 0.2cm,clip]{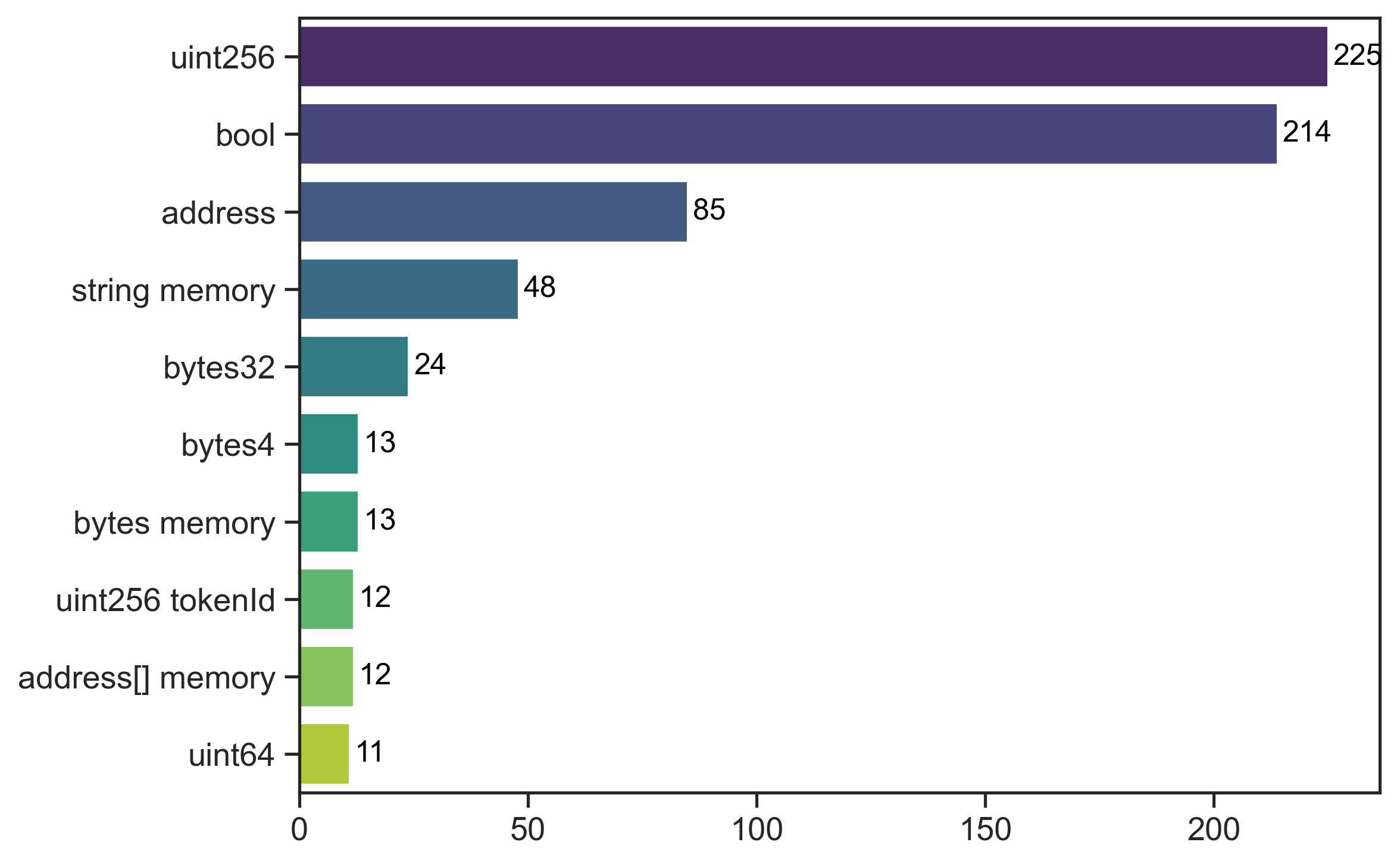}
}
\hfill
\subfigure[Input / Return Parameter Distribution (Occurrences \textgreater~ 5)]{
\label{fig:input_output}
\includegraphics[width=0.31\linewidth]{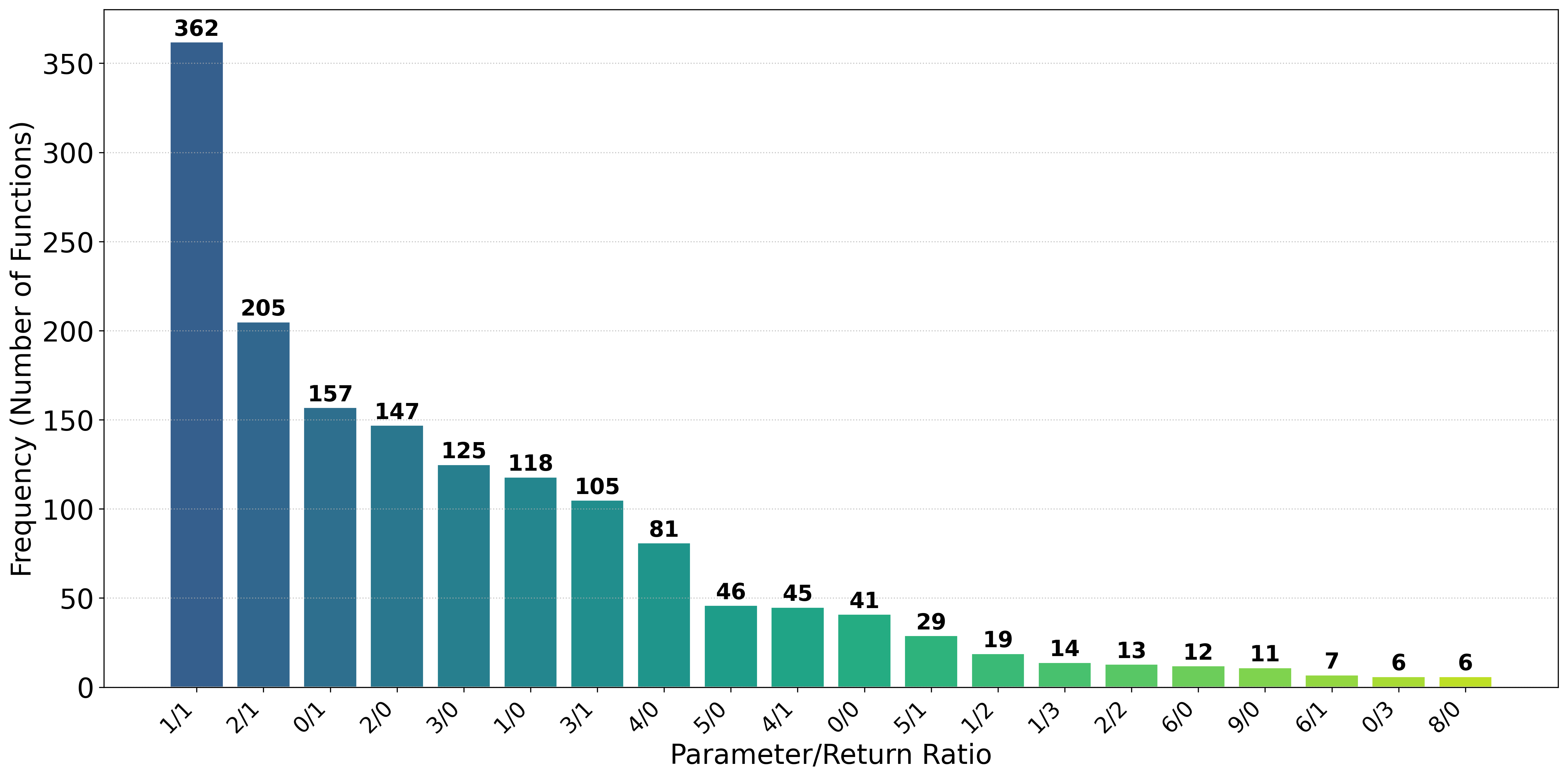}
}

\caption{\textbf{Commonly used parameter types in NFT-related ERC interfaces.} Subfigure (a) shows the top input argument types; (b) presents the most frequent return types; and (c) aggregates input/output parameter categories with more than five occurrences. Together, these reveal strong reliance on basic data types like `address`, `uint256`, and `bool`, reflecting the low-level composability patterns of NFT interfaces.}
\label{fig:combined_param_types}
\end{figure*}

\smallskip
\noindent\textbf{Input and output parameter landscape.}
\textcolor{teal}{Figs.~\ref{fig:input_types}--\ref{fig:output_types}} present the top 10 most frequent parameter types for both inputs and outputs, revealing key patterns in interface design. The input parameters are dominated by \texttt{uint256 tokenId} (381 instances) and its naming variants such as \texttt{\_tokenId}, \texttt{id}, and \texttt{amount}, all of which reflect the centrality of token-level operations. Address-type arguments like \texttt{to}, \texttt{from}, \texttt{owner}, and \texttt{collection} are also prevalent.

On the return side, \texttt{uint256} and \texttt{bool} stand out as the most common types, often used for status checking, identifier retrieval, or internal value exposure. \texttt{address} and \texttt{string memory} appear frequently as return types for ownership verification and metadata presentation. The presence of types such as \texttt{bytes4}, \texttt{bytes32}, and \texttt{address[] memory} suggest support for low-level interface detection, batch operations, and array-based returns, especially in more advanced or composite standards.

\smallskip
\noindent\textbf{Input and return parameter distribution.}
To further explore interface design norms, \textcolor{teal}{Fig.~\ref{fig:input_output}} categorizes functions based on the number of parameters and return values. The most common signature is the \texttt{1/1} pattern (362 times), which often denotes getter-style queries—such as retrieving token metadata or ownership. These read-only operations are simple, efficient, and composable with off-chain systems.

Other dominant patterns include \texttt{2/1} (205), \texttt{0/1} (157), and \texttt{2/0} (147), corresponding to common transfer, query, or approval functions. More complex patterns like \texttt{3/0} (125) and \texttt{3/1} (105) are also notable, supporting batch operations, advanced permissions, or programmable transfers. Long-tail patterns with more than four parameters are rare, usually indicating specialized control flows.

\begin{figure}[!htbp]
    \centering
    \includegraphics[width=\linewidth]{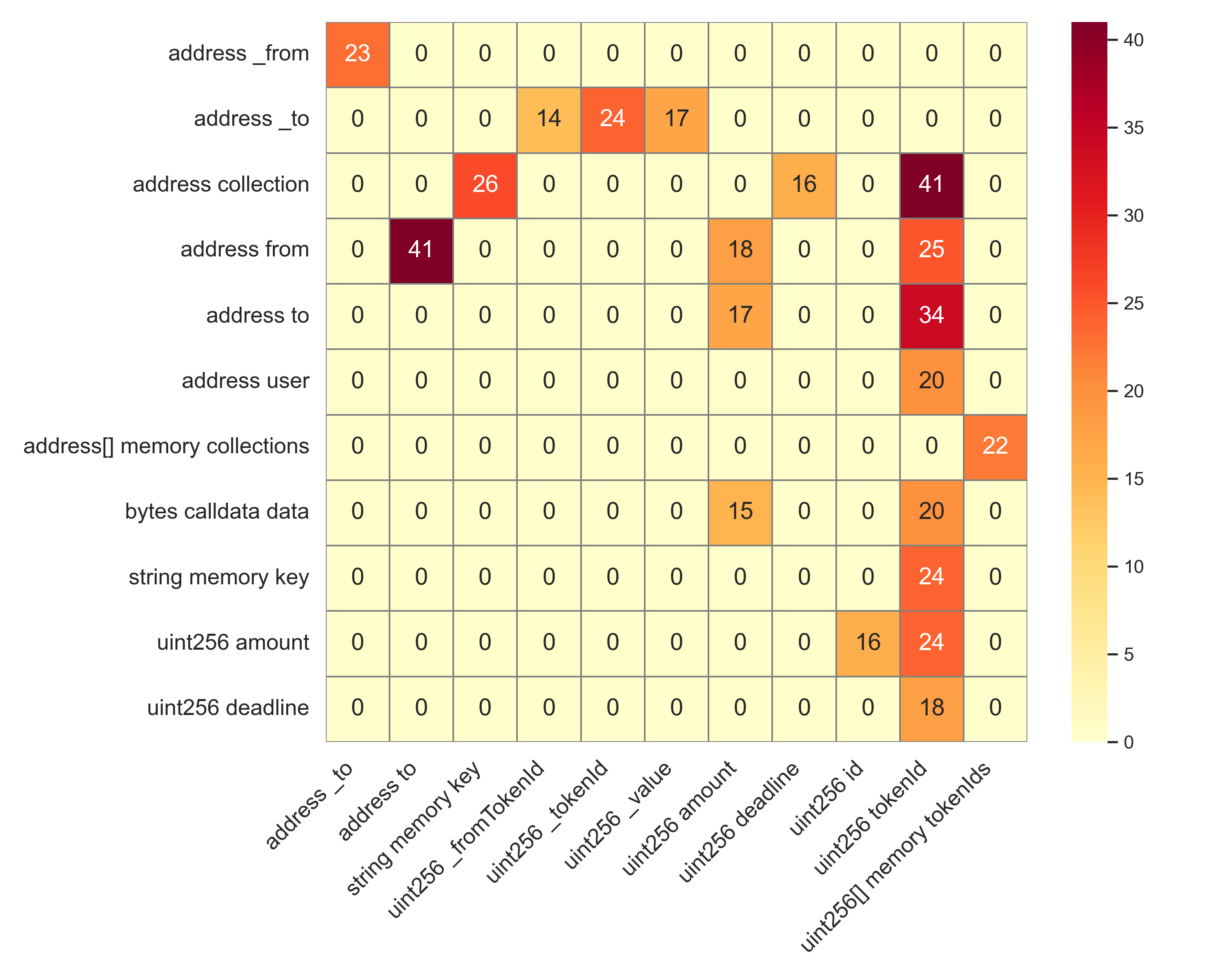}
    \caption{\textbf{Pairwise co-occurrence patterns among input parameter types in NFT-related ERC standards.} Each cell represents the number of functions where the two parameter types appear together, with darker colors indicating higher frequency.}
    \label{fig:input_pairs}
\end{figure}

\smallskip
\noindent\textbf{Co-occurrence patterns among input parameter types.}
We analyze their pairwise co-occurrence across all functions, as illustrated in \textcolor{teal}{Fig.~\ref{fig:input_pairs}}. Each cell in the heatmap quantifies the number of functions that simultaneously utilize two specific parameter types, while the intensity of the color encodes the strength of the association. This analysis reveals both expected and non-trivial patterns of parameter usage that characterize NFT interface design.

At the highest frequency, combinations of address-based parameters dominate, particularly pairs such as \texttt{(address from, address to)} (41 times) and \texttt{(address collection, uint256 tokenId)} (41 times). These pairings highlight the transactional nature of NFT operations, where functions commonly specify both a source and a destination address, or a target collection and an associated token identifier, to complete transfer and approval logic.

A second tier of frequent combinations involves numeric parameters such as \texttt{(address to, uint256 amount)} or \texttt{(address from, uint256 deadline)}, which appear in token distribution or time-sensitive operations. These signatures suggest the presence of extensions beyond simple ownership transfer, supporting functionalities like escrow, batch airdrops, or conditional transfers governed by expiration rules.

Notably, mixed-type patterns involving strings or data payloads, such as \texttt{(bytes calldata data, uint256 id)} or \texttt{(string memory key, uint256 tokenId)}, occur with lower frequency. These functions correspond to specialized contract features, such as off-chain data references, metadata assignment, or authorization flows.

\begin{center}
\tcbset{
    enhanced,
    boxrule=0pt,
    fonttitle=\bfseries
}
\begin{tcolorbox}[
    lifted shadow={1mm}{-2mm}{3mm}{0.1mm}{black!50!white}
]
\textbf{\text{Insight-\ding{205}:}} NFT standards exhibit a highly structured parameter design, dominated by identity-related primitives (\texttt{tokenId}, \texttt{address}) for both input and output types. Pairwise analysis reveals recurring combinations of address and numeric parameters, reflecting the dual need to specify \textit{who} and \textit{what} in token operations.
\end{tcolorbox}
\end{center}

\begin{figure}[b]
    \centering
    \includegraphics[width=\linewidth]{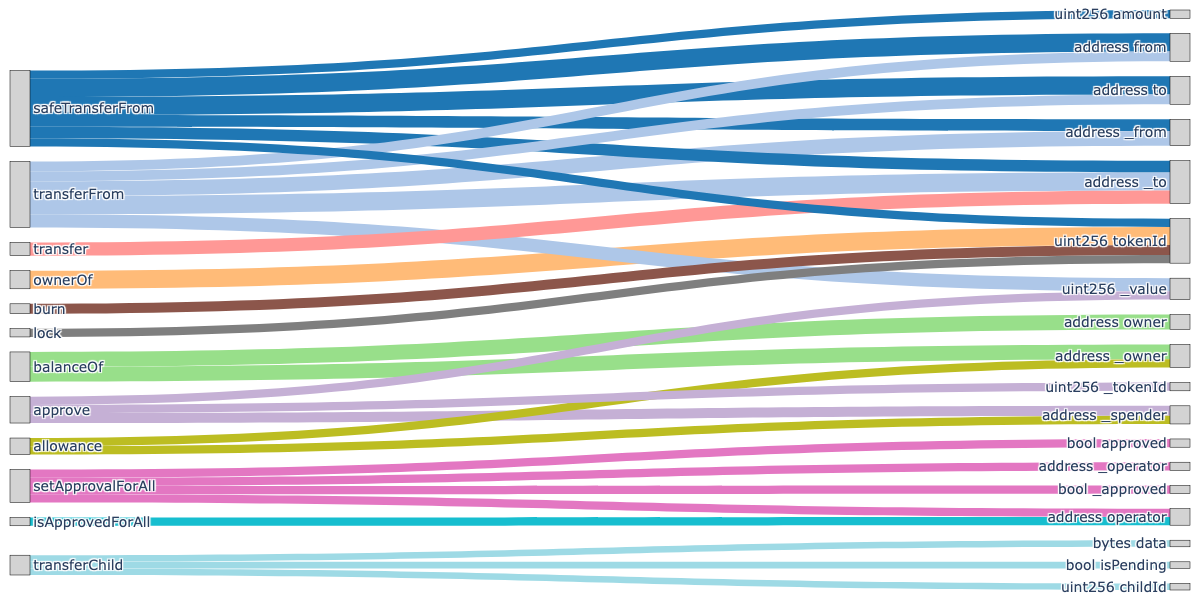}
    \caption{Mapping between NFT function types and their associated input parameter types.}
    \label{fig:function_param_sankey}
\end{figure}

\smallskip
\noindent\textbf{Mapping between function types and parameter types.}
To characterize the design conventions of NFT standards, we examined how different function signatures are associated with specific parameter types. \textcolor{teal}{Fig.~\ref{fig:function_param_sankey}} visualizes this mapping using a Sankey diagram, where flows connect defined functions on the left with their input parameter types on the right. The thickness of each link corresponds to the number of standards implementing the respective function–parameter pairing.

A clear concentration of flows emerges around transfer-related functions. In particular, \texttt{safeTransferFrom} and \texttt{transferFrom} dominate the interaction space, each consistently paired with address-based arguments (\texttt{from}, \texttt{to}, \texttt{\_from}, \texttt{\_to}) and numeric identifiers (\texttt{tokenId}, \texttt{amount}). This pattern reflects the central operational requirement of NFTs: specifying both the source and destination addresses, as well as the unique token instance or quantity involved in each transaction. The high-frequency pairing of these functions with multiple address parameters further illustrates the emphasis on secure routing of asset transfers.

Beyond transfers, several other functions exhibit stable parameter associations. For example, \texttt{approve} and \texttt{setApprovalForAll} are closely linked with permission-oriented arguments such as \texttt{address spender}, \texttt{address operator}, and boolean flags (\texttt{approved}). These signatures capture delegated rights management and batch approvals, which are essential for enabling marketplace listings and third-party custodial operations. Similarly, introspective functions like \texttt{balanceOf} and \texttt{ownerOf} are typically paired with a single \texttt{address} or \texttt{tokenId} parameter, emphasizing the read-heavy operations that underpin asset verification.

\begin{center}
\tcbset{
    enhanced,
    boxrule=0pt,
    fonttitle=\bfseries
}
\begin{tcolorbox}[
    lifted shadow={1mm}{-2mm}{3mm}{0.1mm}{black!50!white}
]
\textbf{\text{Insight-\ding{206}:}} NFT function–parameter mappings are dominated by \texttt{transfer} and \texttt{approval} operations, consistently combining address-based arguments with unique token identifiers. This is a design principle for explicit ownership routing and secure right delegation.
\end{tcolorbox}
\end{center}

\subsection{RQ5: Contributor–Standard Relationships in NFT EIPs}
\label{sec-rq5}

To illuminate the human and organizational dynamics behind NFT EIPs, we examine the contributor–EIP ecosystem from a socio–geographic and participation perspective.

\smallskip
\noindent\textbf{Prevalence of anonymity among NFT-EIP contributors.}
 Authors of EIPs are expected to engage in open, peer-reviewed technical discussions; however, in practice, the blockchain community exhibits a strong culture of privacy preservation. Contributors often adopt pseudonymous identities, relying on handles or aliases rather than verifiable personal information.

 To quantify the degree of anonymity among NFT standard contributors, we analyzed a dataset (\texttt{NFT\_author\_infos.csv}) consisting of 568 author–EIP associations, covering 445 unique authors involved in drafting NFT-related standards. Each author entry was heuristically classified as either:
(1) \textit{Real Name}, when the identifier plausibly corresponds to a verifiable individual (e.g., proper-case personal names); or
(2) \textit{Pseudonym}, when the identifier appears to be an alias, online handle, or unidentifiable nickname (e.g., single-word tags, numeric strings, or unconventional casing).

\begin{figure}[!t]
    \centering
    \begin{minipage}[b]{0.48\columnwidth}
        \centering
        \includegraphics[width=\linewidth]{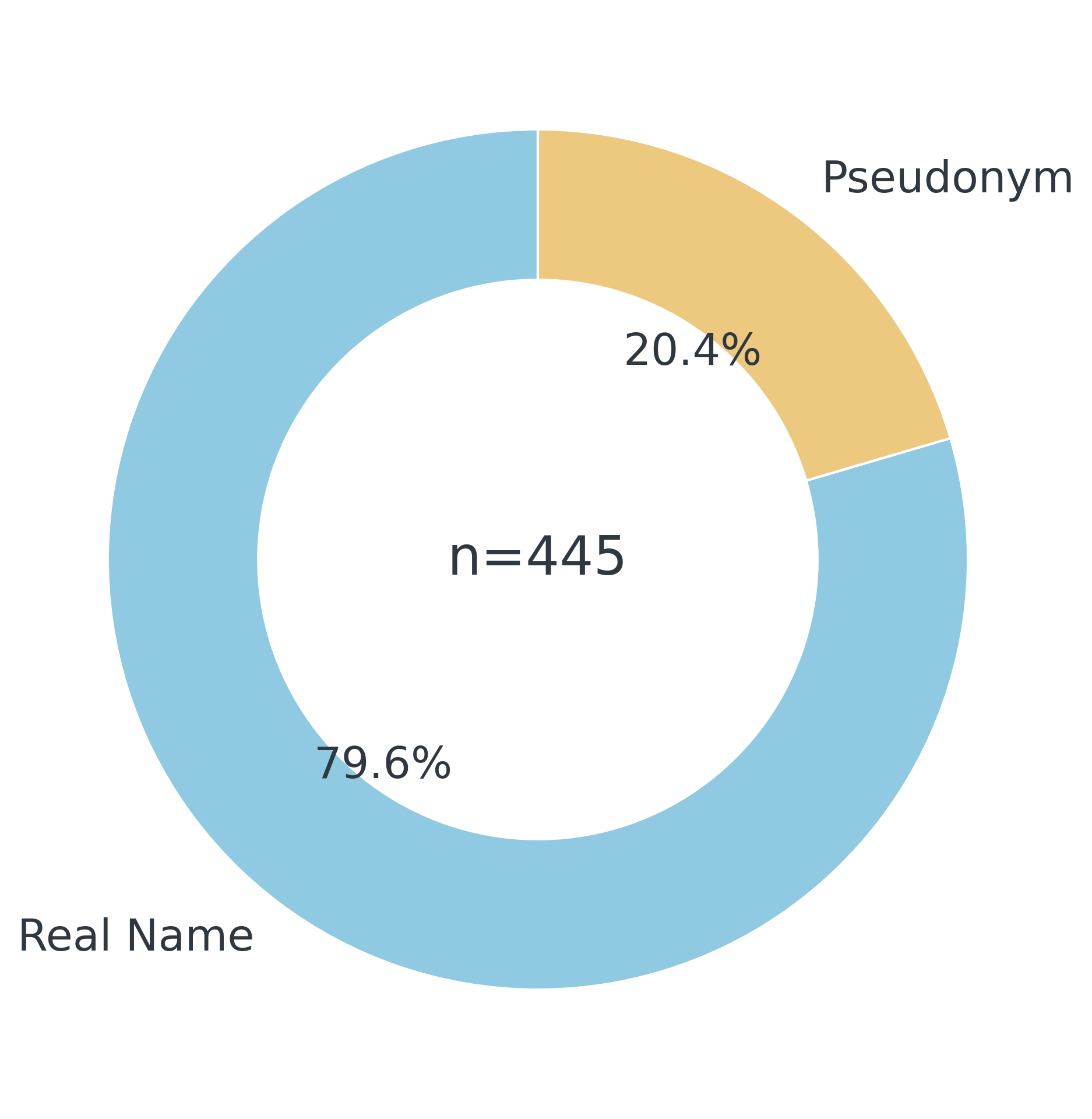}
        \caption*{(a) Unique authors}
    \end{minipage}\hfill
    \begin{minipage}[b]{0.48\columnwidth}
        \centering
        \includegraphics[width=\linewidth]{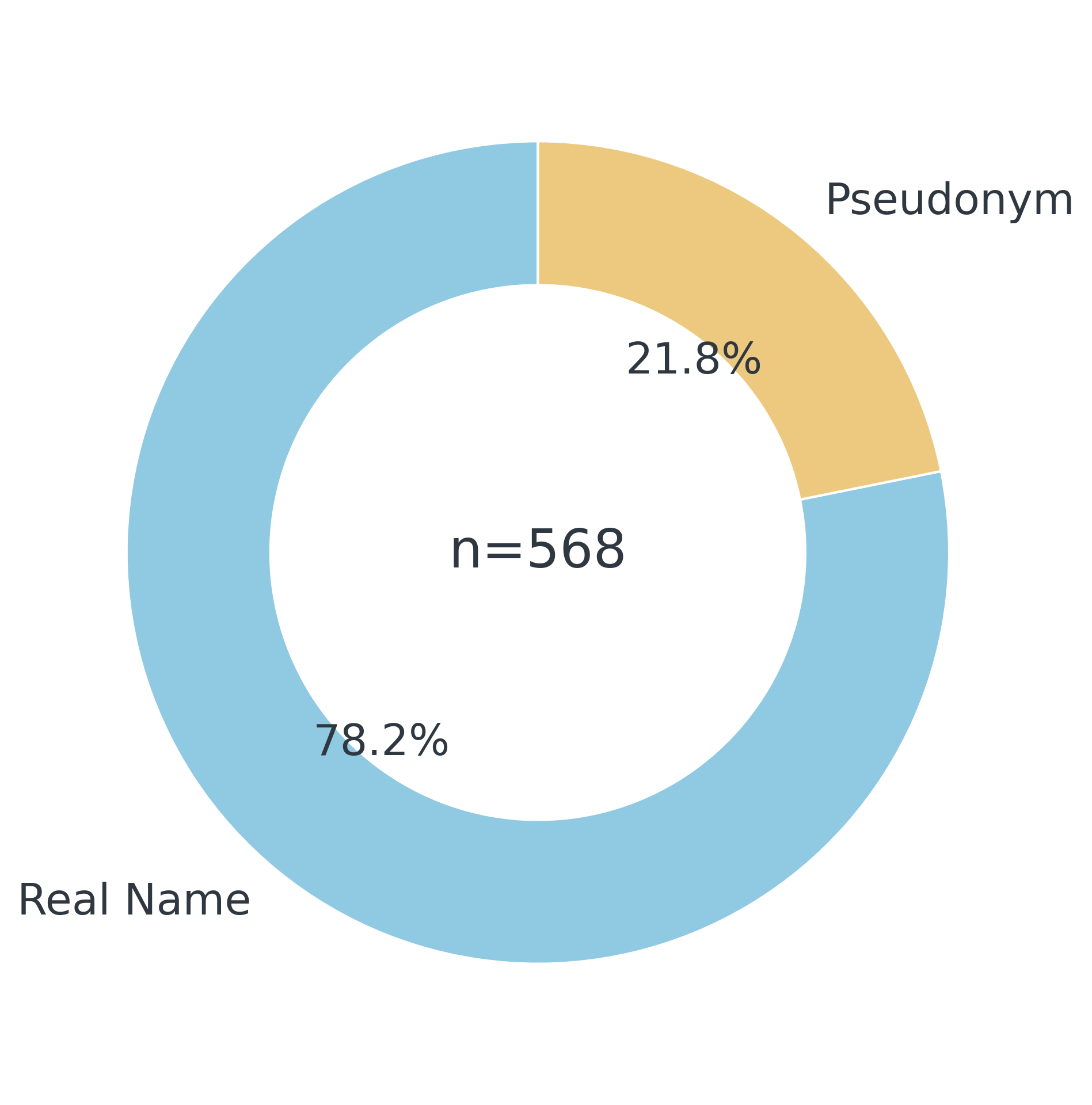}
        \caption*{(b) Author--EIP contributions}
    \end{minipage}
    \caption{\textbf{Prevalence of anonymity or pseudonymity among NFT-EIP contributors.} (a) considers unique authors, while (b) counts contributions, including repeated submissions by the same author.}
    \label{fig:anonymity}
\end{figure}

The analysis was conducted from two complementary perspectives. At the \textbf{contribution level}, 78.2\% of all recorded contributions were made under real names, while 21.8\% were pseudonymous. At the \textbf{unique author level}, 79.6\% of distinct contributors disclosed real names, with 20.4\% relying on pseudonyms. The results are visualized in \textcolor{teal}{Fig.~\ref{fig:anonymity}}, highlighting that while the majority of participation in NFT standardization is attributable to identifiable contributors, a significant minority prefers anonymity or pseudonymity, consistent with broader patterns in decentralized open-source development.

\begin{figure}[!t]
    \centering
     \includegraphics[width=\linewidth]{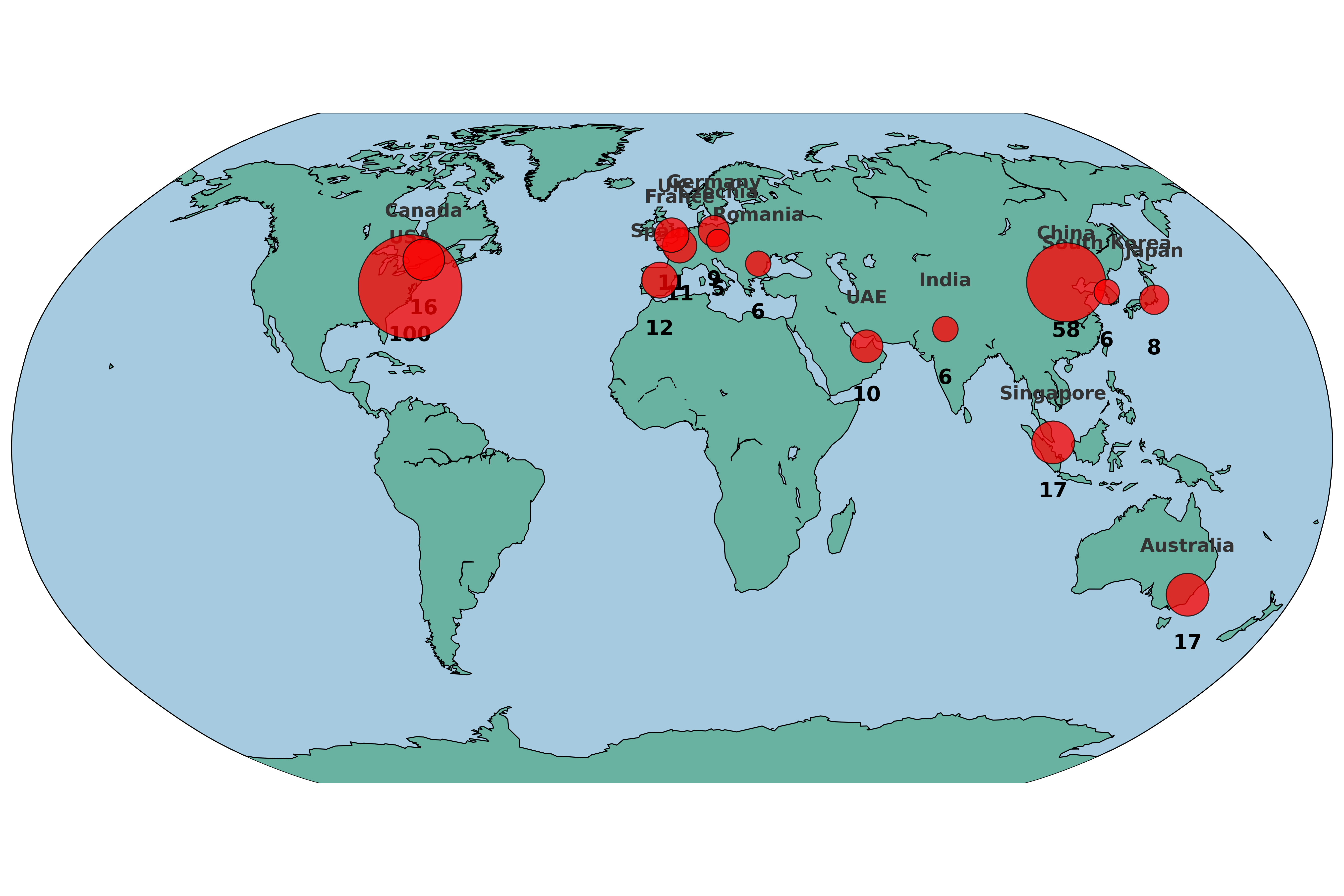}
     \vspace{-0.4in}
    \caption{Top 15 Countries/Regions by Number of Authors}
    \label{fig:bubble_map}
\end{figure}

\smallskip
\noindent\textbf{Geographic distribution of NFT-EIP contributors.}
The development of Ethereum standards, including NFT-related proposals, is inherently global. Contributors originate from diverse jurisdictions, reflecting the open and borderless nature of blockchain communities. To quantify geographic diversity, we analyzed the self-reported or publicly available location information of NFT-EIP authors. Out of 445 unique contributors, 107 (24.2\%) could not be reliably geolocated and were excluded from the country-level distribution analysis. The remaining authors span at least 30 distinct countries or regions, with the top 15 represented in \textcolor{teal}{Fig.~\ref{fig:bubble_map}}.

The United States accounts for the largest share of contributors, representing 22.6\% of identifiable authors. This aligns with its dominant role in global blockchain research and development, hosting major protocol labs, and open-source communities that drive Ethereum ecosystem innovation. China ranks second with 13.0\%, underscoring its vibrant developer base and longstanding engagement with NFT infrastructure projects, despite domestic regulatory fluctuations.

A secondary cluster of active contributors is found in jurisdictions with strong blockchain-friendly ecosystems, including Singapore and Australia (3.8\% each), Canada (3.6\%), and several European countries such as Spain (2.7\%), France (2.5\%), the United Kingdom (2.5\%), and Germany (2.0\%). Other regions, including the UAE, Japan, Romania, India, South Korea, and Czechia, make smaller but notable contributions to the standardization process, reflecting niche participation from individual researchers and localized NFT initiatives.

\begin{figure}[t]
    \centering
    \subfigure[Authors per EIP]{\label{fig:authors_per_EIP}
        \includegraphics[width=0.48\linewidth]{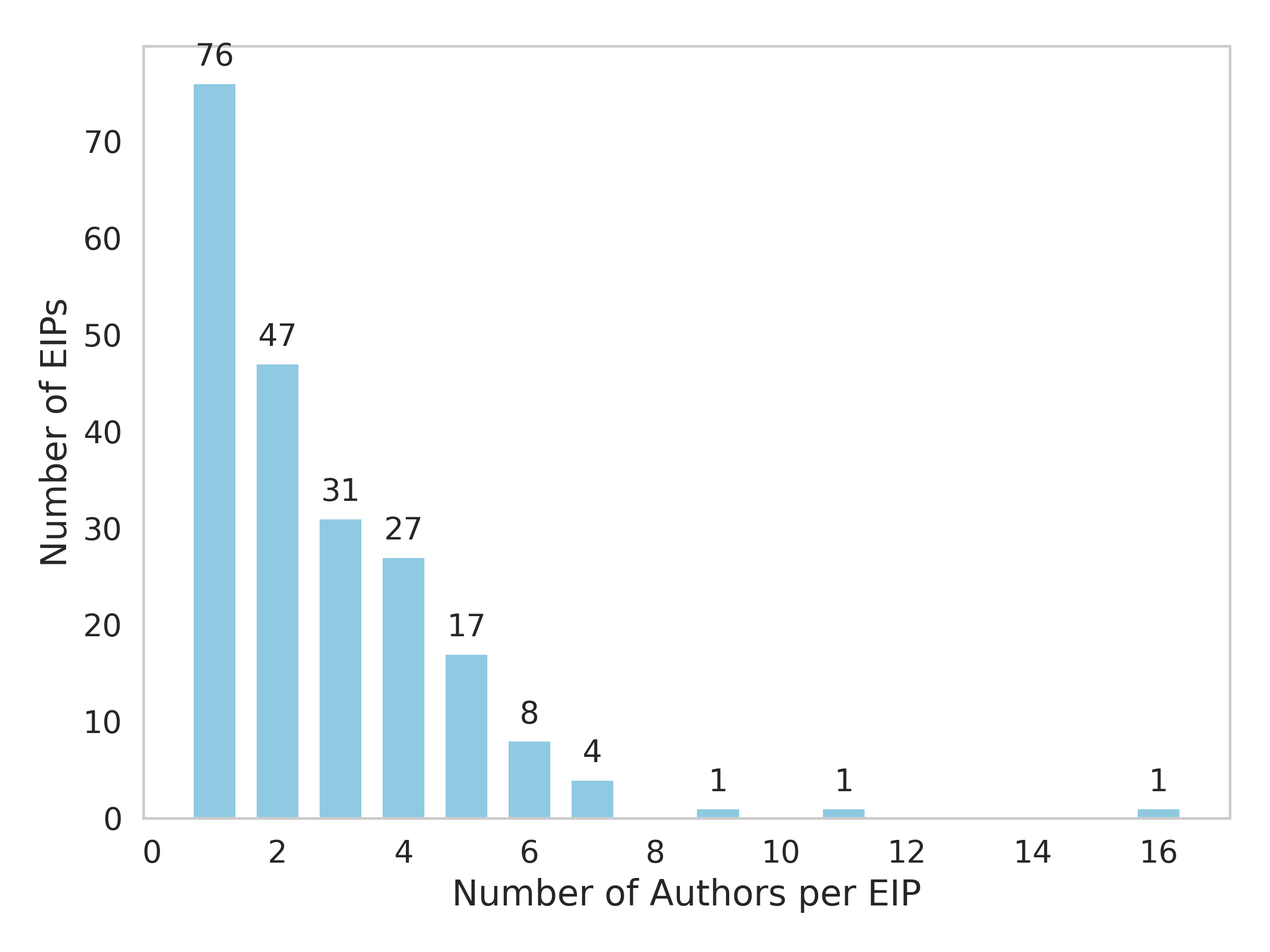}}
    \subfigure[Standards per Author]{\label{fig:standards_per_author}
        \includegraphics[width=0.48\linewidth]{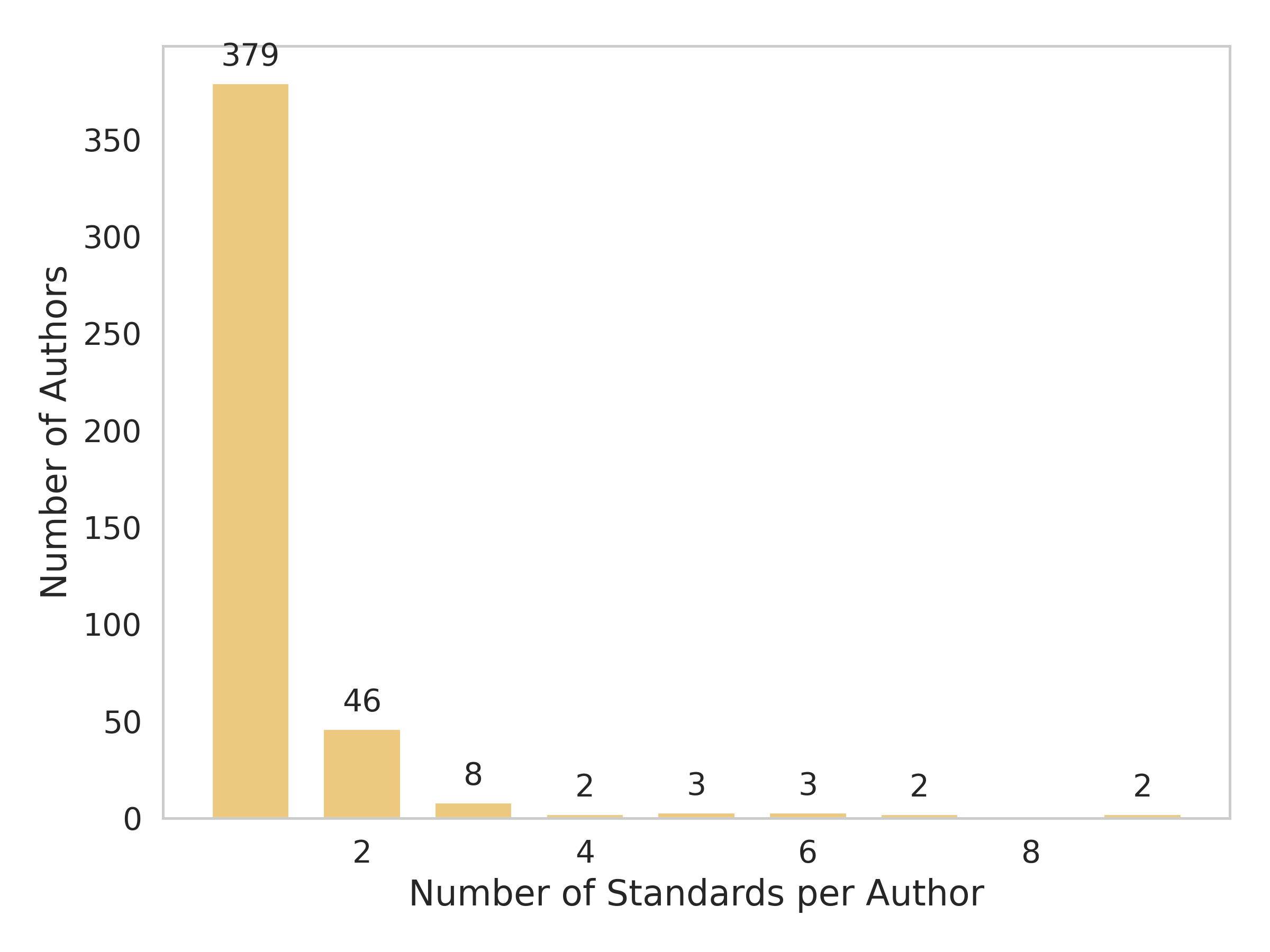}}
    \caption{\textbf{Collaboration patterns in NFT-EIP authorship.} (a) Distribution of authors per standard. (b) Distribution of standards contributed to per author. Both plots highlight a highly skewed participation structure, dominated by single-author EIPs and a small subset of recurring contributors.}
    \label{fig:author_distributions}
\end{figure}

\smallskip
\noindent\textbf{Collaboration patterns in NFT standard authorship.}
To investigate collaboration dynamics in the standardization process, we analyzed the distribution of the number of authors per NFT-EIP and the number of standards to which individual authors contributed. The results, visualized in \textcolor{teal}{Fig.~\ref{fig:author_distributions}}, highlight a strong skew toward single-author proposals and a limited set of contributors driving multiple EIPs.

The first distribution shows that the majority of NFT standards are authored by one or two individuals, with 76 proposals (out of 213) involving a single author and 47 proposals involving two authors. Only a minority of EIPs exhibit broader collaboration, with very few proposals having more than five contributors and a single outlier engaging sixteen authors. This pattern suggests that NFT standardization largely follows a model of small, focused author teams rather than large, committee-driven development.

Conversely, the second distribution reveals that most authors have contributed to only one NFT-based EIP, with 379 authors linked to a single proposal. A small fraction of highly active contributors are involved in two or more standards, and only a handful exceed four contributions. This long-tail pattern indicates that while a large pool of participants engages in standardization sporadically, sustained contributions across multiple EIPs are driven by a relatively small core group of experts or organizations.

\smallskip
\noindent\textbf{Community engagement signals.}  
The resulting analysis reveals \textit{(i) a highly skewed engagement distribution:} the median NFT-related discussion attracted only 6 replies and 5 unique participants, yet a small subset of proposals generated sustained, large-scale debate. For example, the EIP-4973 (\textit{Account-bound Tokens~\cite{daub2022eip4973}}) thread accumulated 100+ replies from over 30 participants between 2022 and 2024, reflecting significant technical contention over identity-binding semantics. Similarly, EIP-6381 (\textit{Public NFT Emote Repository~\cite{eip6381_discussion}}) drew 19 replies despite being relatively recent, indicating concentrated interest in novel user-interaction primitives. \textit{(ii) Engagement longevity also varied markedly:} while 68\% of threads were active for less than one month, several high-profile proposals remained under discussion for over a year, often experiencing bursts of renewed activity following specification updates. These patterns suggest that community attention is not evenly distributed across standards but instead concentrates around proposals introducing fundamentally new technical primitives and touching on sensitive governance.

\begin{center}
\tcbset{
    enhanced,
    boxrule=0pt,
    fonttitle=\bfseries
}
\begin{tcolorbox}[
    lifted shadow={1mm}{-2mm}{3mm}{0.1mm}{black!50!white}
]
\textbf{\text{Insight-\ding{207}:}} NFT standardization shows a skewed participation pattern: authorship is dominated by single-author proposals and few recurring contributors, while community engagement is concentrated on a handful of technically novel or governance-sensitive proposals. 
\end{tcolorbox}
\end{center}

\smallskip
\noindent\textbf{Organizational involvement in NFT standard.}
To complement our author-level analysis, we examine the role of organizations in shaping NFT-related EIPs. This analysis considers both the breadth of organizational contributions and their longitudinal patterns.

\begin{figure*}[t]
  \centering
  \begin{minipage}[t]{0.245\textwidth}
    \centering
    \includegraphics[width=\linewidth]{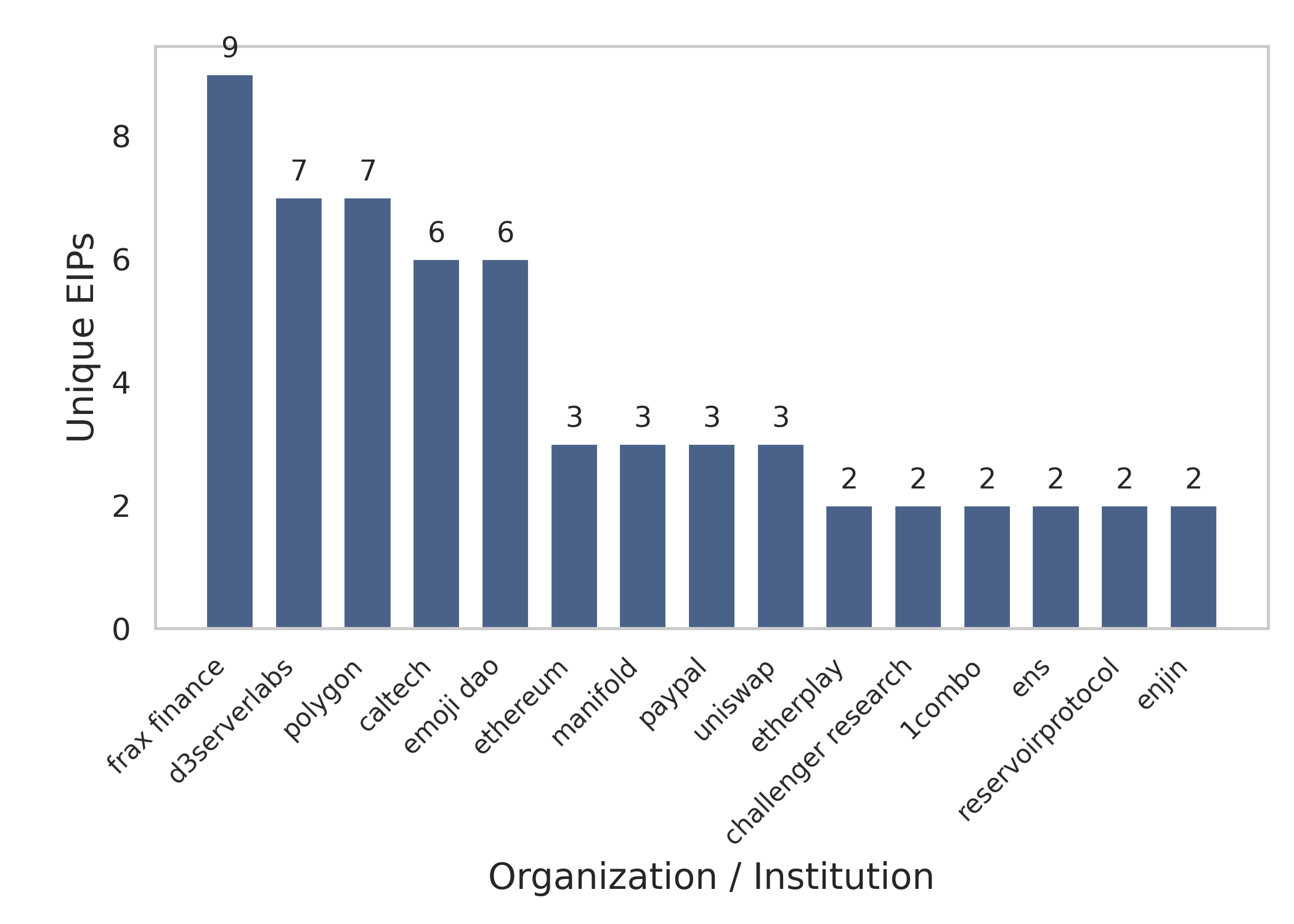}
    \caption*{(a) Top orgs by unique EIPs (excl. Unspecified)}
  \end{minipage}\hfill
  \begin{minipage}[t]{0.245\textwidth}
    \centering
    \includegraphics[width=\linewidth]{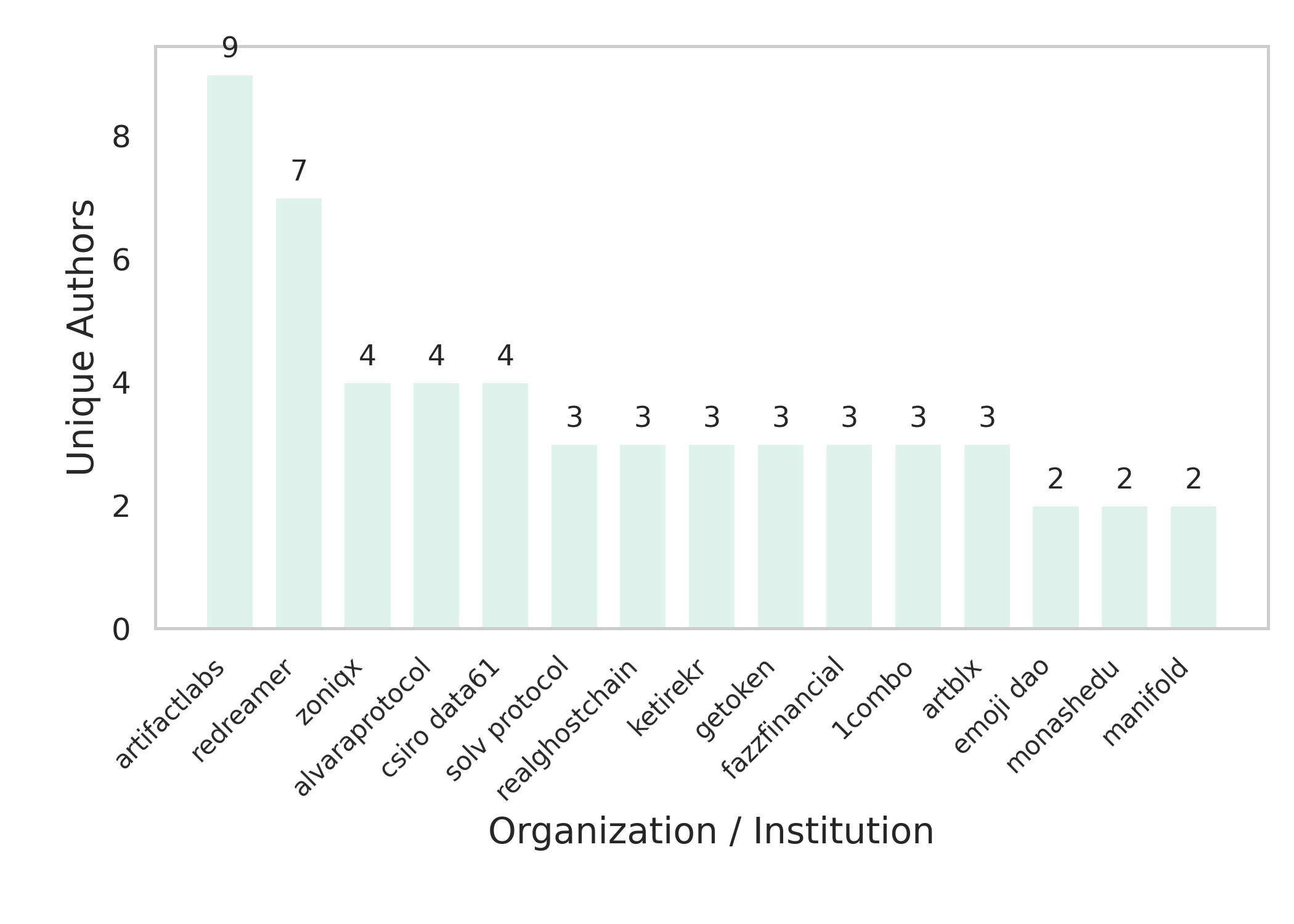}
    \caption*{(b) Top orgs by unique authors (excl. Unspecified)}
  \end{minipage}\hfill
  \begin{minipage}[t]{0.245\textwidth}
    \centering
    \includegraphics[width=\linewidth]{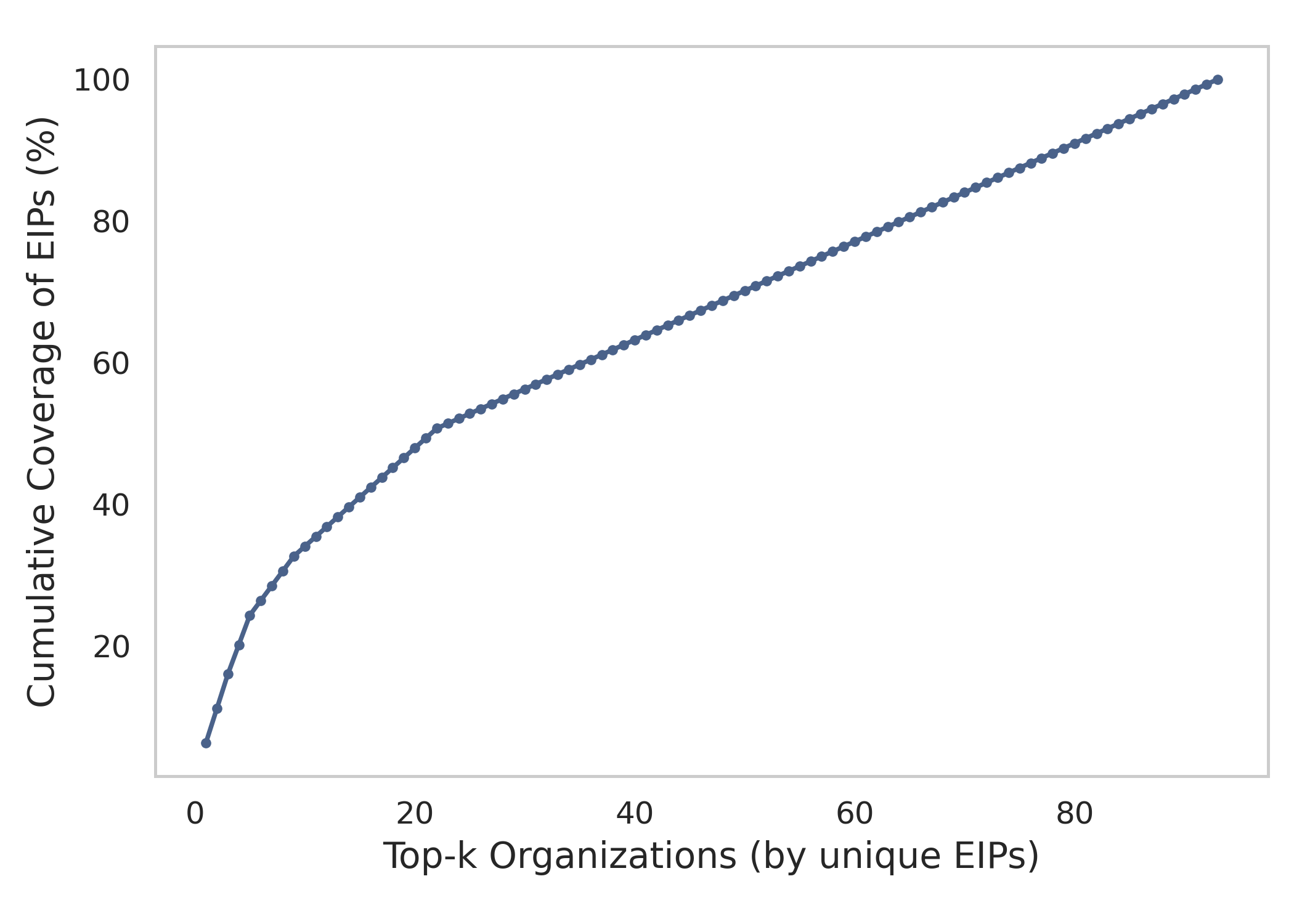}
    \caption*{(c) Pareto coverage of EIPs by orgs}
  \end{minipage}\hfill
  \begin{minipage}[t]{0.245\textwidth}
    \centering
    \includegraphics[width=\linewidth]{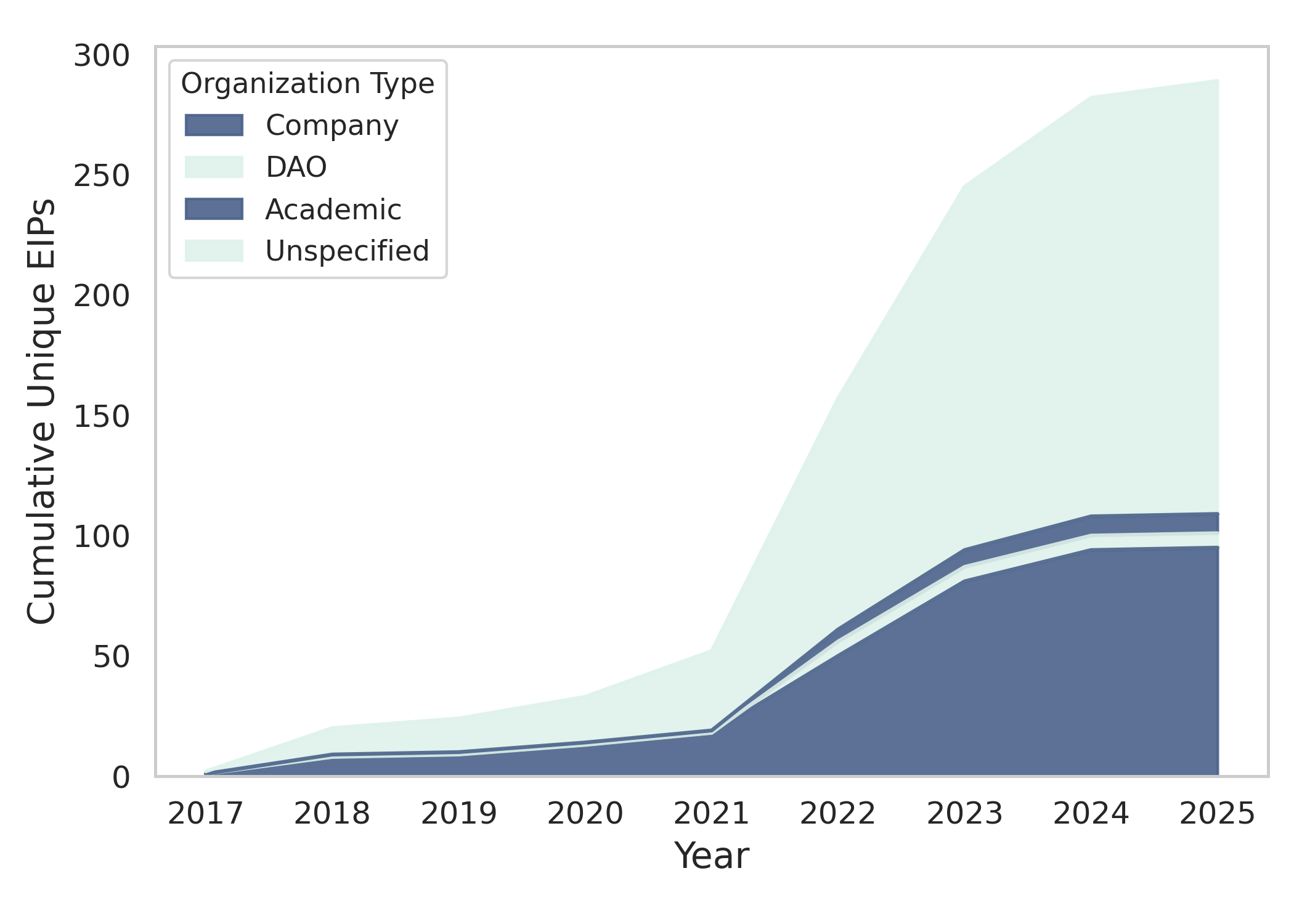}
    \caption*{(d) Cumulative EIPs by org type over years}
  \end{minipage}

  \caption{\textbf{Organizational contributions to NFT-EIPs:} rankings, concentration, and temporal trends.}
  \label{fig:nft_eip_orgs}
\end{figure*}

\textcolor{teal}{Fig.~\ref{fig:nft_eip_orgs}(a)} ranks organizations by the number of distinct NFT EIPs they have contributed to. While entities such as Emoji DAO (11 proposals), Artifactlabs (9), Frax Finance (9), and Polygon (7) emerge as prominent institutional participants, more than 65\% of proposals originate from authors without declared affiliations, classified as \textit{Unspecified/Independent}. This dominance of independent contributors highlights the grassroots nature of NFT standard development, where individual developers remain central to standard-setting activities.

\textcolor{teal}{Fig.~\ref{fig:nft_eip_orgs}(b)} focuses on distinct authors within each organization. Polygon leads with six identified contributors, followed by Emoji DAO and Frax Finance with four to five authors each, suggesting collaborative team-based approaches to drafting proposals. However, the \textit{Unspecified/Independent} group contains over 250 unique authors, representing more than half of all identified contributors. Many of these authors appear only once in the dataset, indicating a broad and decentralized base of occasional participants driving standardization efforts outside formal affiliations.

The Pareto chart in \textcolor{teal}{Fig.~\ref{fig:nft_eip_orgs}(c)} illustrates the cumulative share of EIPs covered by organizations. The top five entities account for roughly 40–45\% of all identified proposals, but the cumulative curve flattens sharply beyond the top ten organizations. This long-tail distribution underscores the fragmented nature of organizational involvement, where a small cluster of entities repeatedly contributes, yet a majority of proposals come from low-frequency participants.

\textcolor{teal}{Fig.~\ref{fig:nft_eip_orgs}(d)} traces the evolution of NFT EIPs by organizational category over time. Company-affiliated contributions grow steadily, surpassing 70 proposals by 2024. DAO-led proposals gain momentum after 2021, reaching approximately 20 cumulative EIPs in recent years. Academic institutions remain marginal, with fewer than 10 contributions overall. Notably, \textit{Unspecified/Independent} authors consistently account for the largest share of proposals across all years, collectively exceeding 100 EIPs. NFT standardization continues to be shaped by individual-driven efforts, with organizations supplementing rather than dominating the process.

\begin{center}
\tcbset{
    enhanced,
    boxrule=0pt,
    fonttitle=\bfseries
}
\begin{tcolorbox}[
    lifted shadow={1mm}{-2mm}{3mm}{0.1mm}{black!50!white}
]
\textbf{\text{Insight-\ding{208}:}} NFT standardization remains a largely decentralized and individual-driven process. While a handful of organizations repeatedly contribute to NFT EIPs, over 65\% of proposals and more than half of unique authors operate without formal affiliations. 
\end{tcolorbox}
\end{center}

\section{Extend Discussion}
\label{sec-extend}
This section provides a broader analysis of NFT-related standards, focusing on the academic research alignment \textbf{(RQ6)} and security risk landscape \textbf{(RQ7)}.

\subsection{RQ6: Academic Research Alignment with NFT Standards}
\label{sec-rq6}

We examine how academic studies engage with NFT-related EIPs, and analyse both the scope of standards referenced and the thematic research threads.

\smallskip
\noindent\textbf{Market and network analyses.}
A substantial body of research investigates NFT market structures, transaction networks, and pricing dynamics, overwhelmingly focusing on ERC‑721 datasets. Nadini et al.~\cite{nadini2021mapping} analyzed 6.1 million trades of 4.7 million NFTs to reveal market specialization and price prediction based on object traits and sales history. Similarly, Barratt et al.~\cite{mekacher2022heterogeneous} and related works observed that rarer ERC‑721 tokens tend to sell at higher prices and exhibit lower trading frequency, confirming the economic significance of rarity traits. Network-focused studies (e.g., Networks of Ethereum NFTs~\cite{casale2021networks}) applied graph analytics to ERC‑721 collections like CryptoPunks and Bored Apes, identifying hubs and community structures in transaction flows. Although ERC‑1155 is mentioned in some gaming studies~\cite{loporchio2024analyzing,tan2024bubble}, its presence is marginal compared to ERC‑721, reflecting the standard’s dominance on marketplaces such as OpenSea.

\smallskip
\noindent\textbf{Comparisons with fungible token networks.}
A few works contrast fungible and non-fungible token behaviors. Loporchio et al.~\cite{loporchio2024comparing} compared ERC‑20 and ERC‑721 transaction graphs, attributing structural differences to ERC‑721’s unique Transfer event that includes a token ID instead of a value parameter. However, such comparisons rarely extend to newer standards.

\smallskip
\noindent\textbf{Metadata permanence and storage.}
Research by Barrington and Merrill~\cite{barrington2022fungibility} and Li et al.~\cite{li2025nfts} highlighted deficiencies in ERC‑721’s metadata persistence, with approximately 45\% of tokens linking to mutable or off-chain resources. These studies reveal the “broken-link” problem, questioning the assumption of immutability for NFTs. Alternative standards or extensions are sometimes proposed to address these issues, but empirical analyses remain narrowly tied to ERC‑721 deployments.

\smallskip
\noindent\textbf{Extended functionality.}
Post‑2021 proposals such as ERC‑ 2981 (royalty enforcement), ‑4907 (rental NFTs), ‑3525 (semi-fungible tokens), and ‑6551 (token-bound accounts) started to appear in academic and policy discussions. EU Blockchain Observatory reports~\cite{ec2021demystifying} emphasize EIP‑2981 for standardizing creator royalties, while research on NFT rentals notes ERC‑4907’s ability to introduce separate user roles and expiry times for temporary rights delegation~\cite{mell2024non}. ERC‑3525 is discussed conceptually as bridging fungibility and non-fungibility, but empirical adoption data is scarce~\cite{li2024empowering}. ERC‑6551 (Token-bound accounts) is highlighted as an extension allowing NFTs to hold other assets and autonomously interact with contracts~\cite{valavstin2024towards}, yet remain at the early stage.

\smallskip
\noindent\textbf{Novel proposals and domain-specific applications.}
Some works~\cite{wang2022erc} propose new standards, such as ERC‑721R for reversible transactions, adding a dispute window and freeze/revert capabilities to improve consumer protection. Others~\cite{li2025nfts, haouari2024secure} suggest referable NFTs (EIP‑5521) enabling directed acyclic graph structures for provenance tracking and NTF fictionalization, respectively. Domain-specific studies~\cite{sibanda2024non, far2022review, wu2022educational, helo2025use} in healthcare, education, gaming, and supply chains frequently invoke NFTs generically, relying implicitly on ERC‑721 without clarifying interoperability or extensions used. This lack of explicit standard references limits their applicability to real-world implementations, where rights enforcement and metadata persistence depend on specific EIPs.


\begin{table*}[!t]
\centering
\caption{Comparison of Key Features and Security Implications of NFT Standards}
\label{tab:erc-comparison-wide}
\resizebox{\textwidth}{!}{%
\begin{tabular}{@{}>{\columncolor{gray!15}}p{3.2cm} p{5.2cm} p{5.2cm} p{5.2cm}@{}}
\toprule
\textbf{Feature} & \textbf{ERC-721 (Unique Asset)} & \textbf{ERC-1155 (Multi-Token)} & \textbf{ERC-6551 (Token-Bound Account)} \\
\midrule
Contract Model &
One contract per token collection or type, providing isolated logic for each series. &
Single contract supports multiple token types (fungible and non-fungible), reducing deployment overhead. &
Extension of ERC-721 where each NFT is linked to its own smart contract wallet (proxy account). \\
\midrule
Transfer Mechanism &
Single transfers via \texttt{safeTransferFrom}; one transaction per token. &
Batch transfers via \texttt{safeBatchTransferFrom} for multiple token types in one call. &
Standard ERC-721 parent transfer; assets inside the TBA are managed separately. \\
\midrule
Gas Efficiency &
\Low{}; every transfer incurs a full transaction cost. &
\High{}; batch operations amortize gas fees across multiple tokens. &
\Variable{}; parent NFT transfer is standard, but TBA management adds extra overhead. \\
\midrule
Metadata Handling &
Each token has a unique \texttt{tokenURI} that references its metadata individually. &
Tokens of the same type can share a URI template with ID-based substitution. &
Inherits ERC-721 structure; no separate metadata standard defined for the TBA itself. \\
\midrule
Vulnerability Vector &
Re-entrancy via \texttt{onERC721Received}; risks from mutable or centralized metadata. &
Re-entrancy via \texttt{onERC1155Received}; additional complexity from batch logic. &
Asset manipulation during sale; issues in account implementation or proxy contract logic. \\
\midrule
Attack Surface &
\RiskLow{Siloed}; a vulnerability typically impacts only one collection. &
\RiskMed{Concentrated}; an exploit can affect all token types under the contract. &
\RiskHigh{Distributed}; risk spans NFT contract, TBA logic, registry, and interacting dApps. \\
\midrule
Audit Complexity &
\LowMed{}; simpler state machines with well-understood patterns. &
\MedHigh{}; batch logic and multi-token interactions increase analysis burden. &
\VeryHigh{}; requires wallet \& proxy mechanism audits with complex external interactions. \\
\bottomrule
\end{tabular}%
}

\vspace{0.3em}
\footnotesize
\raggedright
\textit{Color legend.} \textbf{Gas Efficiency}: \High{} = better (lower cost), \Medium{} = moderate, \Low{} = worse (higher cost), \Variable{} = depends on usage. \quad
\textbf{Audit Complexity}: \Low{} / \LowMed{} = easier, \MedHigh{} = harder, \VeryHigh{} = most demanding. 
\end{table*}

\subsection{RQ7: Security Risk Landscape of NFT Standards}
\label{sec-rq7}

This section characterizes the principal security risks inherent to fundamental and emerging NFT standards (cf. Table~\ref{tab:erc-comparison-wide}).

\smallskip
\noindent\textbf{ERC-721.}
ERC-721 defines a minimal yet expressive interface to manage unique assets, mapping each \texttt{tokenId} to a single owner and supporting core functions such as \texttt{ownerOf}, \texttt{balanceOf}, and secure transfers via \texttt{safeTransferFrom}. An optional metadata extension links each token to off-chain descriptive information via a \texttt{tokenURI}. Despite its simplicity, ERC-721 exhibits recurring classes of vulnerabilities:

\begin{itemize}
    \item \textit{Re-entrancy in safe transfers.} The external call to \texttt{onERC721Received} can be exploited if internal state updates are not completed before execution, enabling malicious callbacks to manipulate contract logic~\cite{cyfrin_reentrancy}.

    \item \textit{Metadata mutability.} Many implementations rely on mutable or centralized URIs for metadata storage~\cite{olympix2024nft}. If external resources become unavailable, the perceived content of an NFT may change or disappear.

    \item \textit{Access control weaknesses.} Poorly protected minting functions have allowed unauthorized token creation, inflating supply and devaluing legitimate assets~\cite{wang2025ai}.

    \item \textit{Gas inefficiency.} ERC-721’s design requires individual transactions for each mint or transfer, leading to high costs in batch operations and making large-scale NFT issuance less practical~\cite{transak2024popular}.
\end{itemize}

\smallskip
\noindent\textbf{ERC-1155.}
The core innovation of ERC-1155 is its two-dimensional state model, mapping \texttt{(owner, tokenId)} pairs to balances. This design supports highly efficient batch operations, allowing multiple tokens to be transferred or minted within a single transaction via functions such as \texttt{safeBatchTransferFrom}. Mandatory receiver checks and the \texttt{onERC1155Received} hook help mitigate risks of token loss when interacting with incompatible contracts. 

While ERC-1155 enhances efficiency, its architectural features create unique risk factors: (i) Because a single ERC-1155 contract may govern a large and diverse set of assets, a single vulnerability can compromise all associated tokens~\cite{rejolut_1155}. In a gaming ecosystem, this could mean the simultaneous theft or manipulation of every in-game item, currency, and land parcel; and (ii) The heterogeneous token model and batch logic raise the cognitive burden on developers and auditors, making it harder to exhaustively test all edge cases and increasing the probability of undetected vulnerabilities~\cite{transak2024popular}.

\smallskip
\noindent\textbf{ERC-6551.}
ERC-6551 extends the NFT model by allowing each ERC-721 token to function as a fully fledged smart contract wallet, known as a Token-Bound Account (TBA). ERC-6551 is built on a modular architecture consisting of: (i) a global Registry for deterministic account creation, (ii) lightweight proxy contracts deployed per NFT to minimize gas costs, and (iii) a shared Account Implementation contract containing execution logic. The key function \texttt{executeCall} enables the NFT owner to perform arbitrary on-chain actions from the TBA’s address, transforming the NFT into a programmable agent. The agent-like nature of TBA broadens the attack surface beyond NFT contract itself with new vectors:

\begin{itemize}
    \item \textit{Asset draining during sales.} A malicious seller may transfer all assets out of a TBA immediately before finalizing a marketplace sale, leaving the buyer with an empty account~\cite{rareskills_erc6551}. This is logically valid on-chain and cannot be mitigated by contract-level checks alone; it requires coordinated safeguards at the protocol level.

    \item \textit{Shared implementation flaws.} A vulnerability in the shared Account implementation could simultaneously compromise thousands of TBAs derived from the same contract~\cite{mundus_eip6551}. Risks include flawed authorization checks in the \texttt{executeCall} function and re-entrancy bugs, each potentially affecting every TBA instance at once.

    \item \textit{Authorization and key management weaknesses.} The linkage between NFT ownership and TBA control is enforced via signature verification logic (e.g., \texttt{isValidSigner})~\cite{goldrush_erc6551}. Errors in this logic, or compromise of the NFT owner’s private key, could lead to total loss of assets stored in the TBA, extending security risks far beyond the NFT itself.
\end{itemize}

\smallskip
\noindent\textbf{Emerging NFT standards.}
Several proposals extend NFT functionality but introduce security and governance risks. ERC-4907~\cite{eip4907} adds time-bound rental rights via a ``user'' role and expiration timestamp, but depends on external escrow or wrapper logic, which may lead to asset loss or double-lending~\cite{debutinfotech_nftrentals}. ERC-4675~\cite{eip4675} supports fractional ownership via vault contracts, creating a central point of failure vulnerable to governance manipulation~\cite{k_fractionalnfts}, and raises unresolved regulatory concerns. ERC-2981~\cite{eip2981} proposes a royalty standard, but lacks on-chain enforcement~\cite{github_erc2981}, relying instead on off-chain marketplaces to voluntarily honor payments~\cite{malone2021exploring}. 

\begin{center}
\tcbset{
enhanced,
boxrule=0pt,
fonttitle=\bfseries
}
\begin{tcolorbox}[
lifted shadow={1mm}{-2mm}{3mm}{0.1mm}{black!50!white}
]
\textbf{\text{Insight-\ding{209}:}} Security risk in NFT standards scales with functional complexity: while ERC‑721/-1155 are comparatively tractable with known mitigations, account‑ and interaction‑centric designs (e.g., ERC‑6551) and newer extensions (e.g., ERC‑4907) introduce enforcement gaps that demand ecosystem‑level safeguards beyond contract correctness.
\end{tcolorbox}
\end{center}

\section{Conclusion}
\label{sec-conclu}
We present the first data-driven analysis of NFT standardization on Ethereum, based on 191 NFT-related EIPs. Our study traces their evolution, functional structures, interdependencies, contributor dynamics, and emerging security concerns. 

We find that core standards dominate adoption and define the baseline for NFT interactions, whereas newer proposals on rentals, programmability, royalties, and fractional ownership remain fragmented. Inheritance analysis indicates limited interface reuse, and contributor profiling reveals a geographically concentrated, largely pseudonymous author base. Community engagement analysis shows a highly skewed distribution: most NFT-related EIP threads attract minimal participation, while a small subset sparks large-scale debate. Our security assessment further uncovers recurring vulnerability patterns, particularly in upgradeability and permission handling.

\normalem
\bibliographystyle{unsrt}
\bibliography{bib.bib}

\input{table}



\end{document}

%% file: table.tex
\begin{table*}[!hbtp]
 \caption{NFT-related ERCs with \textit{Final} Status (Updated on Aug. 2025)}
 \label{tab-final-ercs-nft}
 \centering
 \renewcommand\arraystretch{1}
 \resizebox{\linewidth}{!}{
 \begin{tabular}{cllll}
    \toprule
    \textbf{EIP-} & \textbf{Title} & \textbf{Main (\textcolor{teal}{new}) functions/events/metadata} & \textbf{Feature} & \textbf{Application} \\
    \midrule

    \multicolumn{5}{l}{\textit{\textbf{(A) Core NFT Standards}}} \\
    \hlhref{https://eips.ethereum.org/EIPS/eip-721}{721}   & Non-Fungible Token Standard & \textcolor{teal}{ownerOf, safeTransferFrom, approve} & Unique on-chain assets & Art, Collectibles \\
    \hlhref{https://eips.ethereum.org/EIPS/eip-1155}{1155} & Multi Token Standard & \textcolor{teal}{safeTransferFrom, batchTransfer} & Mixed FT/NFT in one contract & Gaming, Assets \\
    \hlhref{https://eips.ethereum.org/EIPS/eip-3525}{3525} & Semi-Fungible Token & \textcolor{teal}{slotOf, \,SlotChanged} & Slot-based semi-fungibility & Finance, Vouchers \\

    \midrule
    \multicolumn{5}{l}{\textit{\textbf{(B) Metadata, URI \& Presentation}}} \\
    \hlhref{https://eips.ethereum.org/EIPS/eip-1046}{1046} & tokenURI Interoperability & \textcolor{teal}{tokenURI schema hint} & Interop hints for metadata & Marketplaces \\
    \hlhref{https://eips.ethereum.org/EIPS/eip-4906}{4906} & ERC-721 Metadata Update Ext. & \textcolor{teal}{MetadataUpdate (event)} & Notify off-chain indexers & Dynamic NFTs \\
    \hlhref{https://eips.ethereum.org/EIPS/eip-5169}{5169} & Client Script URI for Token Contracts & \textcolor{teal}{clientScriptURI} & Off-chain/scripted rendering & Dynamic UI \\
    \hlhref{https://eips.ethereum.org/EIPS/eip-5219}{5219} & Contract Resource Requests & \textcolor{teal}{resourceURI} & Linked on/off-chain resources & Media, Metadata \\
    \hlhref{https://eips.ethereum.org/EIPS/eip-5625}{5625} & NFT Metadata JSON Schema dStorage Ext. & \textcolor{teal}{dStorage schema} & Decentralized metadata schema & IPFS/Arweave \\
    \hlhref{https://eips.ethereum.org/EIPS/eip-7160}{7160} & ERC-721 Multi-Metadata Extension & \textcolor{teal}{tokenURIs, pinTokenURI} & Multiple URIs per token & Views/Files \\

    \midrule
    \multicolumn{5}{l}{\textit{\textbf{(C) Royalties \& Creator Provenance}}} \\
    \hlhref{https://eips.ethereum.org/EIPS/eip-2981}{2981} & NFT Royalty Standard & \textcolor{teal}{royaltyInfo} & Standardized royalty discovery & Marketplaces \\
    \hlhref{https://eips.ethereum.org/EIPS/eip-4910}{4910} & Royalty Bearing NFTs & \textcolor{teal}{RoyaltyAccount, payout} & On-chain royalty account model & Creator economy \\
    \hlhref{https://eips.ethereum.org/EIPS/eip-5375}{5375} & NFT Author Information and Consent & \textcolor{teal}{authorInfo} & Authorship \& consent meta & IP, Attribution \\

    \midrule
    \multicolumn{5}{l}{\textit{\textbf{(D) Rental, Roles \& Delegation}}} \\
    \hlhref{https://eips.ethereum.org/EIPS/eip-4907}{4907} & Rental NFT (ERC-721 Ext.) & \textcolor{teal}{setUser, userExpires} & Time-bounded user role & Rentals \\
    \hlhref{https://eips.ethereum.org/EIPS/eip-5006}{5006} & NFT User Extension & \textcolor{teal}{UserRecord} & Secondary user semantics & Rentals, Access \\
    \hlhref{https://eips.ethereum.org/EIPS/eip-5007}{5007} & ERC-721 Time Extension & \textcolor{teal}{startTime, endTime} & Time windows for usage & Lending/Access \\
    \hlhref{https://eips.ethereum.org/EIPS/eip-5380}{5380} & ERC-721 Entitlement Extension & \textcolor{teal}{entitle, entitlementOf} & Fine-grained entitlements & Leasing, Sub-licenses \\
    \hlhref{https://eips.ethereum.org/EIPS/eip-6147}{6147} & Guard of NFT/SBT & \textcolor{teal}{changeGuard, transferAndRemove} & Guard role for transfers & SBT/NFT management \\
    \hlhref{https://eips.ethereum.org/EIPS/eip-7432}{7432} & Non-Fungible Token Roles & \textcolor{teal}{grantRoleFrom, roleExpirationDate} & Expirable role assignments & Rentals, Access \\

    \midrule
    \multicolumn{5}{l}{\textit{\textbf{(E) Composability, Nesting \& Multi-asset}}} \\
    \hlhref{https://eips.ethereum.org/EIPS/eip-2309}{2309} & ERC-721 Consecutive Transfer & \textcolor{teal}{ConsecutiveTransfer} & Efficient batch mint event & Large collections \\
    \hlhref{https://eips.ethereum.org/EIPS/eip-5606}{5606} & Multiverse NFTs & \textcolor{teal}{bundle, delegateTokens} & One-to-many asset mapping & Games/Metaverse \\
    \hlhref{https://eips.ethereum.org/EIPS/eip-5773}{5773} & Context-Dependent Multi-Asset Tokens & \textcolor{teal}{acceptAsset, setPriority} & Multi-asset per token & Skins/Views \\
    \hlhref{https://eips.ethereum.org/EIPS/eip-6059}{6059} & Parent-Governed Nestable NFTs & \textcolor{teal}{DirectOwner, Child} & Parent/child nesting & Files, Scenes \\
    \hlhref{https://eips.ethereum.org/EIPS/eip-6150}{6150} & Hierarchical NFTs & \textcolor{teal}{parentOf, childrenOf} & Cross-level hierarchy & File trees \\
    \hlhref{https://eips.ethereum.org/EIPS/eip-6220}{6220} & Composable NFTs (Equippable) & \textcolor{teal}{equip, unequip} & Parts/equipment system & Avatars, Items \\
    \hlhref{https://eips.ethereum.org/EIPS/eip-7401}{7401} & Parent-Governed NFTs Nesting & \textcolor{teal}{nest ops} & Alternative nesting spec & Composability \\

    \midrule
    \multicolumn{5}{l}{\textit{\textbf{(F) Social Graphs, Linking \& Referencing}}} \\
    \hlhref{https://eips.ethereum.org/EIPS/eip-5023}{5023} & Shareable NFT & \textcolor{teal}{share} & Co-ownership/usage signal & Collaboration \\
    \hlhref{https://eips.ethereum.org/EIPS/eip-5489}{5489} & NFT Hyperlink Extension & \textcolor{teal}{authorizeSlotTo, revokeAuthorization} & Hyperlink slots & Cross-collection refs \\
    \hlhref{https://eips.ethereum.org/EIPS/eip-5521}{5521} & Referable NFT & \textcolor{teal}{referringOf, setNodeReferring} & DAG-style referencing & Derivative works \\
    \hlhref{https://eips.ethereum.org/EIPS/eip-6381}{6381} & Public NFT Emote Repository & \textcolor{teal}{emote, emoteCountOf} & Social reactions registry & Social UX \\
    \hlhref{https://eips.ethereum.org/EIPS/eip-7409}{7409} & Public NFT Emote Repository & \textcolor{teal}{emote repo (alt)} & Alternative emote spec & Social \\

    \midrule
    \multicolumn{5}{l}{\textit{\textbf{(G) Soulbound \& Identity NFTs}}} \\
    \hlhref{https://eips.ethereum.org/EIPS/eip-5192}{5192} & Minimal Soulbound NFTs & \textcolor{teal}{locked} & Non-transferable flag & Credentials, Badges \\
    \hlhref{https://eips.ethereum.org/EIPS/eip-5484}{5484} & Consensual Soulbound Tokens & \textcolor{teal}{issueWithConsent} & Consent-bound SBT & Identity \\
    \hlhref{https://eips.ethereum.org/EIPS/eip-6239}{6239} & Semantic Soulbound Tokens & \textcolor{teal}{semantic schema} & Rich semantics for SBT & CV/Profiles \\
    \hlhref{https://eips.ethereum.org/EIPS/eip-7231}{7231} & Identity-aggregated NFT & \textcolor{teal}{setIdentitiesRoot, verifyBinding} & Web2/Web3 bindings & Identity graph \\

    \midrule
    \multicolumn{5}{l}{\textit{\textbf{(H) Authorization, Transferability \& Security}}} \\
    \hlhref{https://eips.ethereum.org/EIPS/eip-5585}{5585} & ERC-721 NFT Authorization & \textcolor{teal}{authorizeUser, transferUserRights} & Cross-user authorization & Access control \\
    \hlhref{https://eips.ethereum.org/EIPS/eip-6066}{6066} & Signature Validation for NFTs & \textcolor{teal}{isValidSignature} & AuthN at NFT layer & E-voting, Access \\
    \hlhref{https://eips.ethereum.org/EIPS/eip-6454}{6454} & Minimal Transferable NFT detection & \textcolor{teal}{isTransferable} & Transferability check & Compliance, UX \\
    \hlhref{https://eips.ethereum.org/EIPS/eip-7007}{7007} & Verifiable AI-Generated Content Token & \textcolor{teal}{evidence, proofs} & AIGC provenance & Media integrity \\

    \midrule
    \multicolumn{5}{l}{\textit{\textbf{(I) Market Protocols \& Commerce}}} \\
    \hlhref{https://eips.ethereum.org/EIPS/eip-4400}{4400} & ERC-721 Consumable Extension & \textcolor{teal}{consume, consumed} & Usage/consumption lifecycle & Tickets, Access \\
    \hlhref{https://eips.ethereum.org/EIPS/eip-6105}{6105} & No Intermediary NFT Trading Protocol & \textcolor{teal}{listItem, buyItem} & P2P marketplace flows & Markets \\
    \hlhref{https://eips.ethereum.org/EIPS/eip-6672}{6672} & Multi-redeemable NFTs & \textcolor{teal}{redeem, getRedemptionIds} & Multiple redemptions & Coupons, Perks \\

    \midrule
    \multicolumn{5}{l}{\textit{\textbf{(J) Phygital, Receipts \& Real-world Links}}} \\
    \hlhref{https://eips.ethereum.org/EIPS/eip-4519}{4519} & NFTs Tied to Physical Assets & \textcolor{teal}{ownerEngagement, userEngagement} & On-chain $\leftrightarrow$ physical link & IoT, Supply chain \\
    \hlhref{https://eips.ethereum.org/EIPS/eip-5570}{5570} & Digital Receipt NFTs & \textcolor{teal}{receipt mint} & Proof-of-purchase & Commerce \\
    \hlhref{https://eips.ethereum.org/EIPS/eip-7578}{7578} & Physical Asset Redemption & \textcolor{teal}{redeem physical} & Off-chain claim & RWAs \\

    \midrule
    \multicolumn{5}{l}{\textit{\textbf{(K) Other NFT-adjacent Primitives}}} \\
    \hlhref{https://eips.ethereum.org/EIPS/eip-5615}{5615} & ERC-1155 Supply Extension & \textcolor{teal}{totalSupply} & Supply discovery & Analytics \\
    \hlhref{https://eips.ethereum.org/EIPS/eip-7631}{7631} & Dual Nature Token Pair & \textcolor{teal}{paired FT/NFT} & FT/NFT pair design & Advanced assets \\
    \hlhref{https://eips.ethereum.org/EIPS/eip-6809}{6809} & Non-Fungible Key Bound Token & \textcolor{teal}{addBindings, approval rules} & Key-bound security for NFTs & Secure transfer \\

    \bottomrule
 \end{tabular}
 }
\end{table*}